\documentstyle[epsfig]{mn2e}

\def\gtsima{$\; \buildrel > \over \sim \;$}
\def\ltsima{$\; \buildrel < \over \sim \;$}
\def\gtrsim{\lower.5ex\hbox{\gtsima}}
\def\lesssim{\lower.5ex\hbox{\ltsima}}

\begin{document}

\title[Impact of metallicity on young SCs]{Impact of metallicity on the evolution of young star clusters}
\author[Mapelli \&{} Bressan]
{M. Mapelli$^{1}$, \&{} A. Bressan$^{2}$
\\
$^1$INAF-Osservatorio Astronomico di Padova, Vicolo dell'Osservatorio 5, I--35122, Padova, Italy; {\tt michela.mapelli@oapd.inaf.it}\\
$^2$Scuola Internazionale Superiore di Studi Avanzati (SISSA), Via Bonomea 265, I--34136, Trieste, Italy
}
\maketitle \vspace {7cm }

  \begin{abstract}
%The mass spectrum of black holes (BHs) originated from the collapse of massive stars is highly uncertain. Massive metal-poor stars might form massive stellar black holes (MSBHs, with mass $>25$ M$_\odot{}$) through direct collapse. In this paper, 
We discuss the results of $N-$body simulations of intermediate-mass  young star clusters (SCs) with three different metallicities ($Z=0.01$, 0.1 and 1 Z$_\odot{}$), including metallicity-dependent stellar evolution recipes and metallicity-dependent prescriptions for stellar winds and remnant formation. The initial half-mass relaxation time of the simulated young SCs ($\sim{}10$ Myr) is comparable to the lifetime of massive stars. We show that mass-loss by stellar winds influences the reversal of core collapse and the expansion of the half-mass radius. In particular, the post-collapse re-expansion of the core is weaker for metal-poor SCs than for metal-rich SCs, because the former lose less mass (through stellar winds) than the latter. As a consequence, the half-mass radius expands faster in metal-poor SCs. The difference in the half-light radius between metal-poor SCs and metal-rich SCs is (up to a factor of two) larger than the difference in the half-mass radius. %This is the result of the intrinsic luminosity and lifetime differences between metal-poor and metal-rich populations.
\end{abstract}

%
%________________________________________________________________

\section{Introduction}\label{sec:intro}
It is well known that metallicity plays an important role in the evolution of massive stars. First, it affects directly the luminosity and effective temperature of massive stars (e.g., Hurley, Pols \&{} Tout 2000; Tumlinson \&{} Shull 2000; Baraffe, Heger \&{} Woosley 2001; Bromm, Kudritzki \&{} Loeb 2001). Secondly, the metallicity has a strong effect on the mass-loss rate by stellar winds (e.g. Kudritzki, Pauldrach \&{} Puls 1987; Leitherer, Robert \&{} Drissen 1992; Maeder 1992; Portinari, Chiosi \&{} Bressan 1998; Kudritzki \&{} Puls 2000; Vink, de Koter \&{} Lamers 2001; Kudritzki 2002; Belkus, Van Bever \&{} Vanbeveren 2007; Pauldrach, Vanbeveren \&{} Hoffmann 2012). This may deeply affect the evolutionary path in the HR diagram up to the formation of the final remnant (e.g. Heger et al. 2003a; Mapelli, Colpi \&{} Zampieri 2009a; Belczynski et al. 2010, hereafter B10). An interesting question is whether, and how much, the above effects can influence the overall evolution of star clusters (SCs).

%The evolution of a star cluster (SC) is strictly connected with the metallicity of its stars, because metallicity affects stellar winds (e.g. Kudritzki, Pauldrach \&{} Puls 1987; Leitherer, Robert \&{} Drissen 1992; Maeder 1992; Kudritzki \&{} Puls 2000; Vink, de Koter \&{} Lamers 2001; Kudritzki 2002; Belkus, Van Bever \&{} Vanbeveren 2007; Pauldrach, Vanbeveren \&{} Hoffmann 2012), stellar luminosities (e.g., Hurley, Pols \&{} Tout 2000) and the formation of remnants (e.g. Heger et al. 2003a; Mapelli et al. 2009a; Belczynski et al. 2010, hereafter B10).

Observations indicate  
%confirm 
that there is a trend with metallicity, at least for globular clusters (GCs). In fact, blue (i.e. generally metal-poor) GCs in the Milky Way and in some nearby galaxies tend to have a larger half-light radius (by $15-20$ per cent) than red (i.e. generally metal-rich) GCs (Kundu \& Whitmore 1998, 2001; Kundu et al. 1999; Puzia et al. 1999; Larsen et al. 2001; Larsen, Forbes \& Brodie 2001; Barmby, Holland \& Huchra 2002; Harris et al. 2002; Jord\'an  2004; Jord\'an et al.  2005, 2009; Harris 2009;  Woodley \&{} G\'omez 2010; Strader et al. 2012). %The hypothesis that this difference can be explained with projection effects (Larsen \&{} Brodie 2003)
No similar studies have been done for young SCs ($<100$ Myr) and for open clusters. 

%Jord\'an et al. (2004) explain the difference in the half-light radii of blue and red GCs as a result of different stellar evolution processes in metal-poor and metal-rich stellar populations. In particular, 
%By means of multimass isotropic Michie-King models, 
%Jord\'an et al. (2004) show that effects of mass segregation combined with the dependence of main sequence (MS) lifetime on metallicity are able to reproduce the observed difference  in the half-light radii of blue and red GCs
Jord\'an (2004) explains the difference in the half-light radii of blue and red GCs as a result of mass segregation combined with the dependence of main sequence (MS) lifetime on metallicity, by means of multimass isotropic Michie-King models. According to Jord\'an  (2004),  the observed difference in the half-light radii does not imply a difference in the half-mass radii.
Recent Monte Carlo (Downing 2012) and $N$-body (Sippel et al. 2012; see also Hurley et al. 2004) simulations of GCs confirm the results by Jord\'an (2004), finding that the difference in the half-light radii arises from the dependence of luminosity and stellar lifetime on metallicity. %Thus, a difference in the half-light radii may not imply a difference in the half-mass radii.
According to Downing (2012), there may be even a difference in the half-mass radii, but only as a consequence of dynamical interactions, mainly due to the presence of massive stellar black holes (BHs). Finally, Schulman, Glebbeek \&{} Sills (2012) ran $N$-body simulations of young intermediate-mass ($10^3-10^4$ M$_\odot$) SCs with different metallicities. They find a $\approx{}10$ per cent difference also in the half-mass radius, between metal-poor SCs and metal-rich SCs.

In this paper, we discuss the results of $N$-body simulations of young intermediate-mass SCs with different metallicity and different recipes for stellar winds and remnant formation. Our aim is to investigate the core collapse and post-collapse evolution of young intermediate-mass SCs, to better understand the interplay between dynamics, metallicity-dependent stellar evolution and formation of stellar remnants.

%Such difference arises from the fact that the relaxation timescale of young intermediate-mass SCs is $\sim{}10-50$ Myr, much shorter than that of GCs

%for young star clusters or open clusters: no studies, small statistics
%Simulations: for globular clusters mainly effect of metal dependence of luminosity, lifetime (Sippel), and also mass of remnants (Downing). For smaller objects different trend

%\section{The basic physics}
\section{The impact of three-body encounters and stellar evolution on core collapse}\label{sec:intro2}
The evaporation of stars from the core of a SC removes part of its kinetic energy (Spitzer 1987). Since a SC  has negative heat capacity, this leads to gravothermal instability and to the collapse of the core (Binney \&{} Tremaine 1987). Core collapse in SCs is reversed mostly by three-body encounters (i.e. close encounters between stars and binaries). In fact, binaries have a energy reservoir (their internal energy, e.g. Binney \&{} Tremaine 1987), which can be exchanged with single stars. In particular, hard binaries (i.e. binaries with binding energy higher than the average kinetic energy of a star in the SC) tend to transfer kinetic energy to single stars as a consequence of three-body encounters (Heggie 1975). The stars that receive this kinetic energy are either ejected from the entire SC or remain in the periphery of the SC.  It is worth mentioning that the more massive a binary is, the higher its expected encounter rate  (e.g. Portegies Zwart 2004). Thus, the most massive binaries in the SC tend to dominate the dynamical evolution of the system (Spitzer 1987; Hurley 2007; Aarseth 2012; Hurley \&{} Shara 2012). %are by far the most important in SC evolution.

It has long been debated whether mass-loss by stellar winds and/or supernovae (SNe) is efficient in affecting core collapse (e.g. Angeletti \&{} Giannone 1977, 1980; Applegate 1986; Chernoff \&{} Shapiro 1987; Chernoff \&{} Weinberg 1990; Hurley et al. 2004; Schulman et al. 2012; Downing 2012; Sippel et al. 2012). In fact, stellar winds and SNe eject mass from a SC, making the central potential well shallower and quenching the onset of gravothermal instability. SCs with a broad mass-range initial mass function (IMF) undergo core collapse on a timescale $t_{\rm cc}\sim{}0.2\,{}t_{\rm h}$ , where $t_{\rm h}$ is the half-mass relaxation timescale (e.g. Portegies Zwart \&{} McMillan 2002). For most GCs, $t_{\rm h}\gtrsim{}1$ Gyr, whereas in young dense SCs $t_{\rm h}\sim{}10-100$ Myr. This means that core collapse in GCs is expected to occur on a timescale (much) longer than the lifetime of massive ($\gtrsim{}20$ M$_\odot$) stars. Thus, the stages of core collapse and post-core collapse are expected to be barely affected by SNe and stellar winds. Instead, the timescale for core collapse in young dense SCs is expected to be of the same order of magnitude as the lifetime of massive stars. Thus, mass-loss by stellar winds and SNe peaks during the epochs of core collapse and post-core collapse. Actually, mass-loss by stellar evolution is expected to delay the core collapse (quenching the gravothermal instability) and/or to reverse more rapidly the core collapse, depending on the interplay between core collapse timescale and massive star lifetime.

%*SCs undergo core collapse on a time-scale sim0.2 two body relaxation time (for a reasonably extended IMF)

%* core collapse is reversed mostly by three-body encounters: interactions binaries - single stars allow to convert a fraction of internal energy into kinetic energy of the cluster

%* there is debate about the role of mass-losses by stellar winds and SN (REFERENCE TO FIRST SIMULATIONS as applegate)

%*important the interplay between core collapse timescale and massive star evolution timescale (portegies zwart? gieles?)
Two further ingredients of this scenario are the dependence of mass-loss on stellar metallicity and the formation of stellar remnants. Stellar winds are suppressed at low metallicity (e.g. Kudritzki, Pauldrach \&{} Puls 1987; Maeder 1992; Kudritzki \&{} Puls 2000; Vink, de Koter \&{} Lamers 2001). Thus, metal-rich SCs are expected to lose more mass by stellar winds than metal-poor ones. 

Massive stars that end their life with mass higher than $\approx{}40$ M$_\odot$ are expected to collapse directly into BHs, with no or faint SN explosion (e.g. Fryer 1999; Fryer \&{} Kalogera 2001). Massive metal-poor stars lose less mass by stellar winds, and thus are more likely to collapse directly into BHs. This mechanism allows to form BHs with mass higher than 25 M$_\odot{}$ (e.g. Mapelli et al. 2009a; B10). If retained inside the SC, these BHs become the most massive objects in the SC after a few tens Myr, dominating the energy budget of three-body encounters.

\section{Method}\label{sec:Method}
The simulations were done using the {\sc starlab}\footnote{\tt http://www.sns.ias.edu/$\sim{}$starlab/} public software environment (Portegies Zwart et al. 2001; see also Portegies Zwart \&{} Verbunt 1996; Nelemans et al. 2001; Anders et al. 2009), which allows to integrate the dynamical evolution of a SC, resolving binaries and three-body encounters. In particular, we used the modified version of {\sc starlab} described in Mapelli et al. (2013, hereafter M13). This version of {\sc starlab} includes recipes for the metallicity dependence of stellar radius, temperature and luminosity, using the polynomial fitting formulae by Hurley, Pols \&{} Tout (2000). It also includes new recipes for mass-loss by winds for MS stars, based on the metallicity-dependent fitting formulae given by Vink et al. (2001; see also B10).

 We added an approximate treatment for luminous blue variable (LBV)  and for Wolf-Rayet (WR) stars. In particular, we assume that a post-MS star becomes a LBV when its luminosity $L$ and radius $R$ satisfy the requirement that $L/{\rm L}_\odot{}>6 \times 10^5$ and $10^{-5}\,{}(R/{\rm R}_\odot{})\,{}(L/{\rm L}_\odot{})^{0.5} >1.0$ (Humphreys \& Davidson 1994). The mass-loss rate by stellar winds for a LBV is then calculated as $\dot{M} = f_{\rm LBV} \times 10^{-4}$ M$_\odot{}$ yr$^{-1}$, where $f_{\rm LBV}=1.5$ (B10).

Naked helium giants coming from stars with zero age MS (ZAMS) mass $m_{\rm ZAMS}>25$ M$_\odot{}$ (e.g., van der Hucht 1991 and references therein) are labelled as WR stars in the new version of the code and undergo a mass-loss rate by stellar winds defined by $\dot{M} = 10^{-13} (L/{\rm L}_\odot{})^{1.5}\,{}({Z/{\rm Z}_\odot})^{\beta{}}$ M$_\odot{}$ yr$^{-1}$, where $\beta{}=0.86$. This formula was first used by B10, and is  a combination of the Hamann \& Koesterke (1998) wind rate estimate (taking into account WR wind clumping) and Vink \& de Koter (2005) wind $Z$-dependence for WR stars.
%, and a very approximate treatment of luminous blue variable (LBV) and Wolf-Rayet (WR) stars (see B10; M13). 
Stellar winds in asymptotic giant branch (AGB) stars are modelled as in the standard version of {\sc starlab} (Portegies Zwart \&{} Verbunt 1996), i.e. the code does not include any recipes for metallicity-dependent stellar winds in AGB stars.

The formation of stellar remnants is implemented as described in M13. In particular, BH masses for various metallicities follow the distribution described in fig. 1 of M13 (see also Fryer et al. 1999; Fryer \&{} Kalogera 2001; B10; Fryer et al. 2012). If the final mass\footnote{We call `final mass', $m_{\rm fin}$, of a star the mass bound to the star immediately before the collapse.} of the progenitor star is $>40$ M$_\odot$, we assume that the SN fails and that the star collapses quietly to a BH. The requirement that $m_{\rm fin}>40$ M$_\odot$ implies that only stars  with ZAMS mass $\gtrsim{}80$ and $\gtrsim{}100$ M$_\odot{}$, can undergo a failed SN at $Z=0.01$ and 0.1 Z$_\odot{}$, respectively. If $m_{\rm fin}\ge{}40$ M$_\odot{}$, the mass of the BH is derived as $m_{\rm BH}=m_{\rm CO}+f_{\rm coll}\,{}(m_{\rm He}+m_{\rm H})$, where $m_{\rm CO}$ is the final mass of the Carbon Oxygen (CO) content of the progenitor, while $m_{\rm He}$ and $m_{\rm H}$ are the residual mass of Helium (He) and of Hydrogen (H), respectively. $f_{\rm coll}$ is the fraction of He and H mass that collapses to the BH in the failed SN scenario.  The value of $f_{\rm coll}$ is uncertain. We assume $f_{\rm coll}=2/3$ to match the maximum values of $m_{\rm BH}$ at low $Z$ derived by B10. In this scenario, BHs with mass  up to $\sim{}80$ M$_\odot$ ($\sim{}40$ M$_\odot$) can form if the metallicity of the progenitor is $Z\sim{}0.01$ Z$_\odot$ ($Z\sim{}0.1$ Z$_\odot$). %This simulation approach is suited for understanding the interplay between dynamical effects (three- and few-body encounters) and metallicity-dependent stellar evolution in SCs.
 BHs that form from quiet collapse are assumed to receive no natal kick (see Fryer et al. 2012). For BHs that form from a SN explosion, the natal kicks were drawn from the same distribution as neutron stars but scaled with the square root of the mass (see M13 for details).

 We assume that the mass lost by stellar winds and SNe is immediately removed from the simulation. This assumption is correct for SN ejecta and also for the winds of massive stars, which are expected to move fast ($\gtrsim{}2000$ km s$^{-1}$ for the O stars, e.g. Muijres et al. 2012; $\gtrsim{}1000$ km s$^{-1}$ for the WR stars, e.g. Vink \&{} De Koter 2005; Martins et al. 2008) with respect to the escape velocity of the simulated SCs ($\lesssim{}10$ km s$^{-1}$). Stellar winds by AGB stars have much smaller velocities ($\approx{}10-20$ km s$^{-1}$, Loup et al. 1993), but still sufficiently high to escape from our simulated SCs. Furthermore, we show in Section~\ref{sec:result} that AGB stars do not play an important role for the results presented in this paper.

\subsection{Initial conditions and simulation grid}
The main properties of the simulated SCs are the same as described in M13. In particular, we focus on intermediate-mass ($M_{\rm TOT}=3000-4000$ M$_\odot{}$) young ($<100$ Myr) SCs. We assume a spherical King profile with central dimensionless potential $W_0=5$ (King 1966), initial core radius $r_{\rm c}=0.4$ pc, concentration $c=\log_{10} (r_{\rm t}/r_{\rm c})=1.03$ (where $r_{\rm t}$ is the tidal radius).  The resulting half-mass radius is $r_{\rm hm}\sim{}0.8-0.9$ pc. The basic SC properties are listed in Table~1.
 %%%%%%%%%%%%%%%%%%%%%%%%%%%%%%% TABLE 1%%%%%%%%%%%%%%%%%%%%%%%%%%%%%%%%%
\begin{table}
\begin{center}
\caption{SC properties in initial conditions.} \leavevmode
\begin{tabular}[!h]{ll}
\hline
Parameter & Values \\
\hline
$W_0$ & 5 \\
$N_\ast{}$ & 5500 \\
$r_{\rm c}$ (pc) & 0.4\\
$c$ & 1.03\\
IMF & Kroupa (2001)\\
$m_{\rm min}$ (M$_\odot{}$) & 0.1\\
$m_{\rm max}$ (M$_\odot{}$) & 150\\
$f_{\rm PB}$ & 0.0, 0.1 \\
$Z\,{}({\rm Z}_\odot{})$ & 0.01, 0.1, 1.0\\
\noalign{\vspace{0.1cm}}
\hline
\end{tabular}
\begin{flushleft}
\footnotesize{$W_0$: central dimensionless potential in the King (1966) model; $N_{\ast}$: number of stars per SC;  $r_{\rm c}$: initial core radius; $c\equiv{}\log{}_{10}{(r_{\rm t}/r_{\rm c})}$: concentration ($r_{\rm t}$ is the initial tidal radius); $m_{\rm min}$ and $m_{\rm max}$: minimum and maximum simulated stellar mass, respectively; $f_{\rm PB}$: fraction of primordial binaries,  defined as the number of primordial binaries in each SC divided by the number of `centres of mass' (CMs) in the SC. In each simulated SC, there are initially 5000 CMs, among which 500 are designated as `binaries' and 4500 are `single stars' (see Downing et al. 2010 for a description of this formalism). Thus, 1000 stars per SC are initially in binaries.}
\end{flushleft}
\end{center}
\end{table}
%%%%%%%%%%%%%%%%%%%%%%%%%%%%%%%%%%%%%%%%%%%%%%%%%%%%%%%%%%%%%%%%%%%%%%%%%%%%%

The initial centres of mass (CMs) of the particles in each simulation are 5000. Each CM corresponds either to a single star or to the CM of a binary. This formalism is commonly used in simulations of SCs including primordial binaries (e.g., Downing et al. 2010). Note that the fraction of primordial binaries $f_{\rm PB}$ is defined as the number of binaries, divided by the total number of CMs. Thus,  $f_{\rm PB}=0.1$ means that there are 500 binaries over 5000 CMs (i.e. 5500 stars).
The single stars and the primary members of a binary follow a Kroupa initial mass function (IMF, Kroupa 2001), with minimum and maximum mass equal to 0.1 and 150 M$_\odot{}$, respectively. 
The masses of the secondary stars ($m_2$) are generated according to a uniform distribution between $0.1\,{}m_1$  and $m_1$ (where $m_1$ is the mass of the primary).  The initial semi-major axis $a$ of a binary is chosen from a distribution $f(a)\propto{}1/a$ (Sigurdsson \&{} Phinney 1993; Portegies Zwart \&{} Verbunt 1996), consistent with the  observations of binary stars in the Solar neighbourhood (e.g. Kraicheva et al. 1978; Duquennoy \&{} Mayor 1991). We generate $a$ between R$_\odot{}$  and $10^5\,{}$R$_\odot{}$, but discarding systems where the distance between the two stars at the pericentre is smaller than the sum of their radii (Portegies Zwart, McMillan \&{} Makino 2007). The initial eccentricity $e$ of a binary is  chosen from a thermal distribution $f(e)=2\,{}e$, in the $0-1$ range (Heggie 1975).

The  half-mass relaxation time for the simulated SCs is $t_{\rm h}\sim{}10\,{}{\rm Myr}\,{}(r_{\rm h}/0.8\,{}{\rm pc})^{3/2}\,(M_{\rm TOT}/3500\,{}{\rm M_\odot})^{1/2}$. Thus, the core collapse time (Portegies Zwart \&{} McMillan 2002) is $t_{\rm cc}\approx{}2-3\,{}{\rm Myr}\,{}(t_{\rm h}/10\,{}{\rm Myr})$. %These timescales indicate that the strongest dynamical evolution of the SC (core collapse and post-core collapse phase) coincides with the lifetime of massive stars ($t\sim{}3$ Myr being the MS lifetime of the most massive stars in the simulations).
We integrate the evolution of the SCs for the first 100 Myr, i.e. the epoch when the interplay between strong dynamical interactions and massive stellar evolution is more important. 
The properties of the grid of simulations are summarized in Table~2. For each SC model, we run a number  $N_{\rm re}$ of single realizations (changing only the random seeds), to filter out the fluctuations associated with each single realization and to get an `average' model. Runs A1, B and C are our fiducial runs (i.e. the models described in M13) and differ only for the metallicity ($Z=0.01$, 0.1 and 1 Z$_\odot{}$, respectively\footnote{In our simulations, we assume Z$_\odot{}=0.019$.}). The runs labelled with A$i$ (where $i=2,3,4,5$) have the same metallicity ($Z=0.01$ Z$_\odot{}$) as our fiducial runs A1, and differ from A1 for other parameters. In particular, (i) in A2 $f_{\rm PB}=0$, (ii) in A3 the maximum allowed BH mass is $m_{\rm BH,\,{}max}=25$ M$_\odot{}$, (iii) in A4 all BHs receive a natal kick  $v_{\rm kick}=10^3$ km s$^{-1}$ (set to an unrealistically high value to eject all BHs from the SC, for comparison with Downing 2012), and (iv) in A5 no mass-loss by stellar winds was implemented.
%%%%%%%%%%%%%%%%%%%%%%%%%%%%%%% TABLE 2%%%%%%%%%%%%%%%%%%%%%%%%%%%%%%%%%
\begin{table}
\begin{center}
\caption{Simulation grid.} \leavevmode
\begin{tabular}[!h]{lllllll}
\hline
Name & $N_{\rm re}$ & Z             & Stellar winds & BH kick & $f_{\rm PB}$ &  $m_{\rm BH,\,{}max}$  \\
     &              &  (Z$_\odot{}$) &               &         &            &  (M$_\odot{}$) \\
\hline
A1   & 100          & 0.01           & YES       & LOW        & 0.1        &  80\\
B    & 100          & 0.1            & YES       & LOW        & 0.1        &  40 \\
C    & 100          & 1.0            & YES       & LOW        & 0.1        &  23 \\
A2   & 100          & 0.01           & YES       & LOW        & 0.0        &  80 \\
A3   & 50           & 0.01           & YES       & LOW        & 0.1        &  25 \\
A4   & 22           & 0.01           & YES       & HIGH       & 0.1        &  80 \\
A5   & 22           & 0.01           & NO        & LOW        & 0.1        & 110 \\
\noalign{\vspace{0.1cm}}
\hline
\end{tabular}
\begin{flushleft}
\footnotesize{Column 1: Name of the set of runs; Column 2, $N_{\rm re}$: number of random realizations per each SC model; Column 3: metallicity; Column 4: YES/NO distinguishes between models in which stellar winds are/are not included; Column 5: LOW/HIGH distinguishes between models where BH natal kicks are described as in M13 (i.e., no kick is assigned to BHs with mass $>25$ M$_\odot{}$; whereas the natal kicks were drawn from the same distribution as neutron stars but scaled with the square root of the mass for  BH masses $\le{}25$ M$_\odot{}$), and models where a natal kick velocity $v_{\rm kick}=10^3$ km s$^{-1}$ was assigned to all BHs (this velocity was purposely set to an unrealistically high value, to eject all BHs from the SC). Column 6, $f_{\rm PB}$: fraction of primordial binaries; Column 7: $m_{\rm  BH,\,{}max}$ is the maximum possible mass for BHs. Runs A1, B, C, A2 and A3 were already presented in M13.}
\end{flushleft}
\end{center}
\end{table}
%%%%%%%%%%%%%%%%%%%%%%%%%%%%%%%%%%%%%%%%%%%%%%%%%%%%%%%%%%%%%%%%%%%%%%%%%%%%%

 The properties of the simulated SCs (total mass, number of stars, core density, core and half-mass radius) are consistent with the properties of observed young intermediate-mass SCs (see e.g. the recent review by Portegies, McMillan \&{} Gieles 2010; see also Hillenbrand \&{} Hartmann 1998; Dias et al. 2002; Portegies Zwart 2004; Pfalzner 2009; Kuhn et al. 2012). Finally, our simulations do not include recipes for the tidal field of the host galaxy. Accounting for the tidal field may increase the fraction of mass lost and even transform the SCs into unbound associations (e.g. Gieles \&{} Portegies Zwart 2011). The effect of tidal fields will be added and discussed in forthcoming papers. In this paper, we decided not to include tidal fields because we want to look at the intrinsic properties of the simulated SCs.
%%%%%%%%%%%%%%%%%%%%%%%%%%%%%%%%%%% FIGURE 1 %%%%%%%%%%%%%%%%%%%%%%%%%%%%%%%%%%
\begin{figure}
\center{{
\epsfig{figure=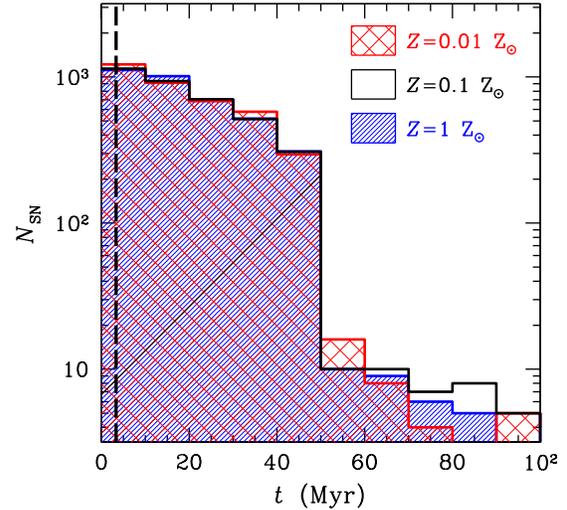,height=7.5cm} 
}}
\caption{\label{fig:fig1}
Number of core-collapse SNe in our simulations as a function of time for $Z=0.01$ Z$_\odot{}$ (cross-hatched red histogram, runs A1), $Z=0.1$ Z$_\odot{}$ (black empty histogram, runs B) and $Z=1$ Z$_\odot{}$ (hatched blue histogram, runs C). Vertical dashed line: time below which SNe are failed for $Z=0.01$ Z$_\odot{}$  and $Z=0.1$ Z$_\odot{}$, in our simulations. In this and in all the figures of this paper, colours are available in the online version.
}
\end{figure}
%%%%%%%%%%%%%%%%%%%%%%%%%%%%%%%%%%%%%%%%%%%%%%%%%%%%%%%%%%%%%%%%%%%%%%%%%%%%%%%

%%%%%%%%%%%%%%%%%%%%%%%%%%%%%%%%%%% FIGURE 2 %%%%%%%%%%%%%%%%%%%%%%%%%%%%%%%%%%
\begin{figure}
\center{{
\epsfig{figure=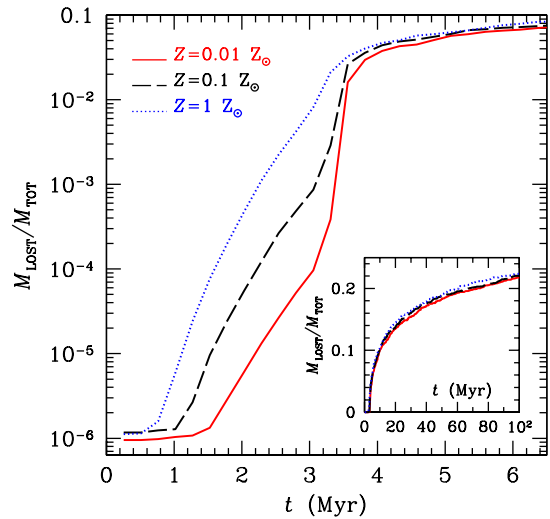,height=7.5cm} %was mloss.eps
}}
\caption{\label{fig:fig2}
 Cumulative mass-loss by stellar winds and SNe ($M_{\rm LOST}$) normalized to the initial total mass of the SC ($M_{\rm TOT}$) as a function of time for the first 6.5 Myr. In the insert: $M_{\rm LOST}/M_{\rm TOT}$ is shown for 100 Myr. Solid red line: A1 ($Z=0.01\,{}$Z$_\odot{}$); dashed black line: B ($Z=0.1\,{}$Z$_\odot{}$); dotted blue line: C ($Z=1\,{}$Z$_\odot{}$). Each line in this Figure is the median value of 100 simulated SCs.
}
\end{figure}
%%%%%%%%%%%%%%%%%%%%%%%%%%%%%%%%%%%%%%%%%%%%%%%%%%%%%%%%%%%%%%%%%%%%%%%%%%%%%%%

\section{Results}\label{sec:result}
The simulated SCs have initial half-mass relaxation time $t_{\rm h}\sim{}10\,{}{\rm Myr}$ and core collapse time $t_{\rm cc}\approx{}3\,{}{\rm Myr}$ (see the previous section). These timescales are of the same order of magnitude as the lifetime of the most massive stars. The lifetime of a 30 M$_\odot{}$ star is $\sim{}6$ Myr, and stellar winds are relatively inefficient for smaller stellar masses (excluding the AGB phase). Thus, the most intense phase of mass-loss by stellar winds coincides with the collapse and re-expansion of the SC core:
%and the peak of core-collapse SNe coincides with the collapse and re-expansion of the SC core. Thus, the effects of mass-loss by stellar winds and SN explosions on core collapse and post-core collapse phase should be maximal in these SCs.
thus, the effects of stellar winds on the structural properties of the SC should be maximal in our simulations.

%The effect of 
 Core-collapse SNe span a longer time-range with respect to stellar winds. Fig.~\ref{fig:fig1} shows that the rate of SNe is maximum at $t\sim{}3-10$ Myr, remains quite high up to $\approx{}50$ Myr,  and drops at $t>50$ Myr. The few  SNe at $t>50$ Myr involve blue straggler stars (BSSs), which are the result of either mass transfer or a merger between two MS stars (e.g. Mapelli et al. 2004, 2006, 2009b).  For $50<t/{\rm Myr}<100$, the number of SN explosions involving BSSs  is $\sim{}0.35$ per SC (regardless of the metallicity), corresponding to a  mass-loss of $\sim{}2.5$ M$_\odot{}$ per SC. Since the total mass-loss in the time interval between 50 and 100 Myr is $\approx{}3.5$ per cent of the initial total mass of the SC ($M_{\rm TOT}$), i.e. $\sim{}100-140$ M$_\odot{}$ per SC, the mass lost through SN explosions of BSSs is negligible.
%The contribution of BSSs to the total mass-loss is negligible, as the number of SN explosions involving BSSs per SC is $\sim{}0.35$ (regardless } 
Thus, after $t\gtrsim{}50$ Myr, the evolution of the core is dominated by three-body interactions and stellar winds of AGB stars.

 For comparison, most GCs have half-mass relaxation timescales and core collapse timescales that are a factor of $\gtrsim{}10$ longer (e.g. Portegies Zwart 2004; Portegies Zwart, McMillan \&{} Gieles 2010), indicating that the peak of stellar wind activity and that of SN explosions ended well before the beginning of the core instability phase.

\subsection{Metallicity dependence}
Fig.~\ref{fig:fig2} shows the cumulative mass lost by the SC (because of both stellar winds and SNe), as a function of time, for runs A1, B and C. The effect of different metallicities is maximum between 1 and 4 Myr, when the mass lost at $Z=1$ Z$_\odot$ is up to 100 and 10 times higher than the  mass lost at $Z=0.01$ Z$_\odot$ and $Z=0.1$ Z$_\odot$, respectively. The mass lost in these early phases of the SC life is a relatively small fraction ($<0.1$) of the total SC mass, but is large when compared to the core mass (which is $\lesssim{}0.1\,{}\,{}M_{\rm TOT}$). As the most massive stars already sank to the core through dynamical friction by the time of core collapse, the effect of early mass-loss is particularly strong in the SC core. For example, at $t=3.3$ Myr the mass-loss is $\sim{}30$ per cent of the core mass at $Z=1$ Z$_\odot{}$, and only $\sim{}1$ per cent at $Z=0.01$ Z$_\odot{}$.

%\subsection{Metallicity dependence}
%AGGIUNGERE PLOT DELLE MASSE
%The fact that the core mass at different metallicity remains different for a time much longer than the lifetime of massive stars implies that mass cannot be re-accreted efficiently by the core. Thus, the effect of metallicity on core evolution is expected to last much longer than the lifetime of very massive stars, before the effect of mass-loss by stellar winds is washed out by other effects.}
%%%%%%%%%%%%%%%%%%%%%%%%%%%%%%%%%%% FIGURE 3 %%%%%%%%%%%%%%%%%%%%%%%%%%%%%%%%%%
\begin{figure}
\center{{
\epsfig{figure=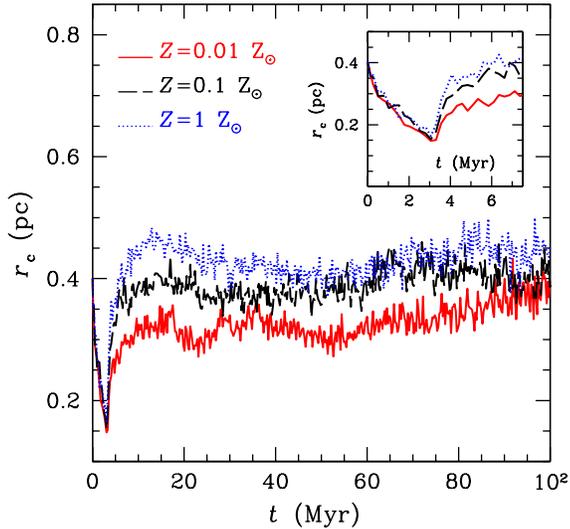,height=7.5cm} %was fig2.eps
}}
\caption{\label{fig:fig3}
Core radius ($r_{\rm c}$) as a function of time for the three considered metallicities: $Z=0.01\,{}$Z$_\odot{}$ (solid red line, A1), $Z=0.1\,{}$Z$_\odot{}$ (dashed black line, B) and $Z=1\,{}$Z$_\odot{}$ (dotted blue line, C). Each line in this Figure is the median value of 100 simulated SCs. In the insert: zoom of the first 7.5 Myr.
}
\end{figure}
%%%%%%%%%%%%%%%%%%%%%%%%%%%%%%%%%%%%%%%%%%%%%%%%%%%%%%%%%%%%%%%%%%%%%%%%%%%%%%%
%%%%%%%%%%%%%%%%%%%%%%%%%%%%%%%%%%% FIGURE 4 %%%%%%%%%%%%%%%%%%%%%%%%%%%%%%%%%%
\begin{figure}
\center{{
\epsfig{figure=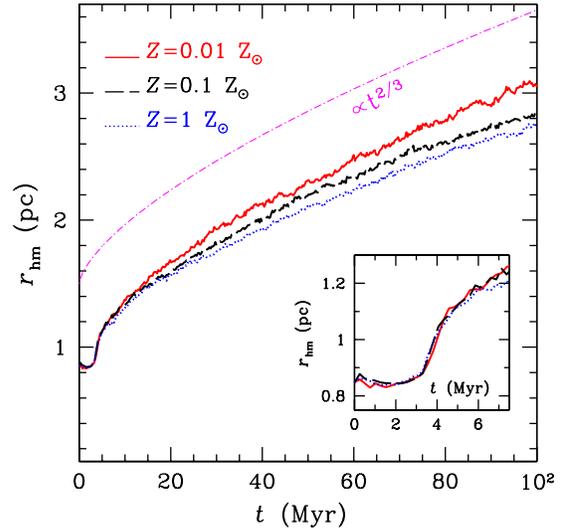,height=7.5cm} %was fig3.eps
}}
\caption{\label{fig:fig4}
Half-mass radius ($r_{\rm hm}$) as a function of time for the three considered metallicities: $Z=0.01\,{}$Z$_\odot{}$ (solid red line, A1), $Z=0.1\,{}$Z$_\odot{}$ (dashed black line,  B) and $Z=1\,{}$Z$_\odot{}$ (dotted blue line,  C). Each of the aforementioned three lines in this Figure is the median value of 100 simulated SCs. The dot-dashed magenta line shows the analytical prediction. In the insert: zoom of the first 7.5 Myr. 
}
\end{figure}
%%%%%%%%%%%%%%%%%%%%%%%%%%%%%%%%%%%%%%%%%%%%%%%%%%%%%%%%%%%%%%%%%%%%%%%%%%%%%%%
Fig.~\ref{fig:fig3} shows the behaviour of the core radius $r_{\rm c}$ as a function of time for runs A1, B and C (i.e. for the three considered metallicities). The collapse is so fast ($t_{\rm cc}\sim{}3$ Myr) that it occurs almost at the same time for all the considered metallicities. %Otherwise, if f $t_{\rm cc}>3$ Myr, we expect that core collapse is delayed in metal-rich star clusters, because of the mass lost in stellar winds.
The effect of metallicity in Fig.~\ref{fig:fig3} appears immediately after the collapse, during the first phase of re-expansion: the core radius in metal-rich SCs expands more than in metal-poor SCs, as mass-loss by stellar winds in metal-rich SCs removes more matter from the core potential well. %After $t\sim{}30$ Myr, corresponding to the MS lifetime of a $\sim{}8$ M$_\odot{}$ star, $r_{\rm c}$ becomes almost the same for all the considered metallicities. At this point, the contribution of core collapse SNe and stellar winds by massive stars ($>20$ M$_\odot{}$) is over: the evolution of the core is completely determined by three-body interactions.
 At $t\approx{}6$ Myr, the mass-loss by stellar winds is nearly over, but the differences among core radii at different metallicity remain almost constant up to $t\sim{}30$ Myr. 

The contribution of SN explosions lasts for a longer time ($t\sim{}50$ Myr, Fig.~\ref{fig:fig1}). In our models, the mass-loss by SN explosions does not depend on metallicity. The only important effect of metallicity on SNe is that failed SNe can take place only at low metallicity ($\lesssim{}0.1$ Z$_\odot{}$). All the failed SNe occur at $t<3.5$ Myr (dashed line in Fig.~\ref{fig:fig1}), as only the most massive stars (see Section~\ref{sec:Method}) can undergo a failed SN. Thus, the occurrence of failed SNe in low-metallicity SCs  enhances the difference between metal-rich and metal-poor SCs, in the early phase of core-collapse reversal.

At times $t>50$ Myr, even the mass-loss by SNe is over: the only process that may affect significantly the later evolution of the core radius is represented by three-body encounters. The differences between metal-rich and metal-poor SCs tend to be quenched at late times. %After $t\sim{}90$ Myr, $r_{\rm c}$ becomes almost the same for all the considered metallicities.
%At this time, the contribution of  stellar winds by massive stars ($>30$ M$_\odot{}$) and even that of core-collapse SNe is over.}
%In fact, the rate of core collapse SNe is maximum at $t\sim{}5$ Myr and drops at $t\sim{}50$ Myr (see Fig.~\ref{fig:fig1}). The few  SNe at $t>50$ Myr involve blue straggler stars that are the result of a stellar merger. Thus, after $t\sim{}50$ Myr, the evolution of the core is completely determined by three-body interactions.}

\begin{figure}
\center{{
\epsfig{figure=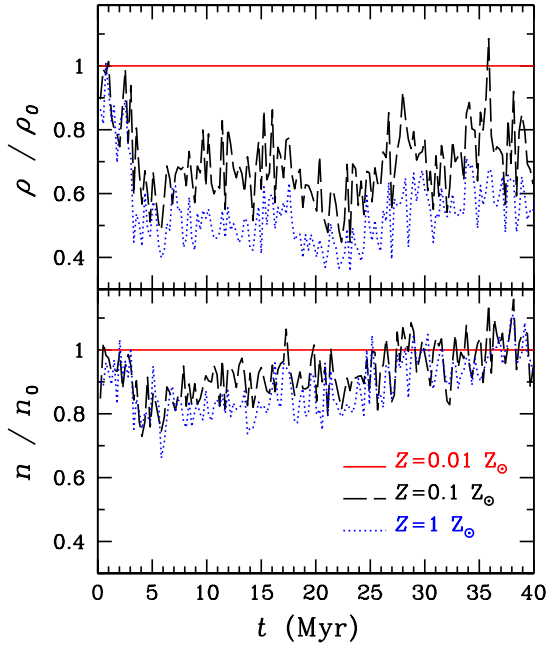,width=7.5cm} %was dens.eps fig3b
}}
\caption{\label{fig:fig5}
 Top panel: core mass density of stars ($\rho{}$) as a function of time for the three considered metallicities: $Z=0.01\,{}$Z$_\odot{}$ (solid red line,  A1), $Z=0.1\,{}$Z$_\odot{}$ (dashed black line,  B) and $Z=1\,{}$Z$_\odot{}$ (dotted blue line,  C). Each line in this Figure is the median value of 100 simulated SCs and is normalized to $\rho{}_0$, i.e. the core mass density of stars in the case of $Z=0.01\,{}$Z$_\odot{}$. Bottom panel: core number density of stars ($n$) as a function of time for the three considered metallicities: $Z=0.01\,{}$Z$_\odot{}$ (solid red line,  A1), $Z=0.1\,{}$Z$_\odot{}$ (dashed black line,  B) and $Z=1\,{}$Z$_\odot{}$ (dotted blue line,  C). Each line in this Figure is the median value of 100 simulated SCs and is normalized to $n_0$, i.e. the core number density of stars in the case of $Z=0.01\,{}$Z$_\odot{}$.
}
\end{figure}

The half-mass radius $r_{\rm hm}$ (Fig.~\ref{fig:fig4}) remains almost constant during core collapse (as expected, e.g. Elson, Hut \&{} Inagaki 1987), while it starts increasing after the reverse of the core collapse. The behaviour of  $r_{\rm hm}$ ($r_{\rm hm}\propto{}t^{2/3}$) is in agreement with simple analytical predictions (e.g. Elson et al. 1987). The post-core collapse value of  $r_{\rm hm}$ for metal-poor SCs is systematically larger than that for metal-rich SCs, in agreement with Schulman et al. (2012). 
%Intuitively, the reason is that core collapse is more dramatic and its reversal is slower for metal-poor SCs. This implies that metal-poor SCs reach higher core densities (see Fig.~\ref{fig:fig5}) in the late core-collapse phase and in the core-collapse reversal. 
% and thus a higher rate of three-body encounters than metal-rich SCs.
 The reason is that the reversal of core-collapse is slower for metal-poor SCs. This implies that metal-poor SCs maintain a higher core density in the late core-collapse phase and in the early core-collapse reversal (see Fig.~\ref{fig:fig5} for the evolution of core stellar density\footnote{The core mass (number) density of stars was approximated as the total mass (number) of stars in the core divided by $r_{\rm c}^3$.}). Since the rate of three-body encounters scales approximately with the stellar mass density (e.g. Sigurdsson \&{} Phinney 1993), metal-poor SCs have a higher rate of three-body encounters than metal-rich SCs. Three-body encounters pump kinetic energy in the SC halo (mainly in the form of stars ejected in the outskirts of the SC), and are responsible for the expansion of $r_{\rm hm}$ (e.g., Elson et al. 1987).

%The reversal of core collapse in metal-poor clusters relies more on stars ejected after three-body encounters than on mass-losses by stellar winds. While the mass ejected by stellar winds is completely lost by the cluster potential well, The stars that undergo three-body encounters may  either be completely ejected  from the SC or remain bound to the SC, but on radial orbits that reach the outskirts of the SC. Thus, the fraction of stars ejected in the outskirts of the SC to reverse core collapse  is higher in metal-poor SCs than in metal-rich SCs.
%%%%%%%%%%%%%%%%%%%%%%%%%%%%%%%%%%% FIGURE 5 %%%%%%%%%%%%%%%%%%%%%%%%%%%%%%%%%%
%%%%%%%%%%%%%%%%%%%%%%%%%%%%%%%%%%%%%%%%%%%%%%%%%%%%%%%%%%%%%%%%%%%%%%%%%%%%%%%

The difference at $t\gtrsim{}40$ Myr between the averaged  $r_{\rm hm}$ of $Z=0.01$ Z$_\odot$ SCs and that of $Z=1$ Z$_\odot$ SCs is $\sim{}10$ per cent, surprisingly similar to the difference observed between red and blue GCs. On the other hand, we stress that the evolution of our simulated young SCs is very different from the evolution of GCs. %%%Thus, this is likely a coincidence. 
%%%%%%%%%%%%%%%%%%%%%%%%%%%%%%%%%%% FIGURE 6 %%%%%%%%%%%%%%%%%%%%%%%%%%%%%%%%%%
\begin{figure}
\center{{
\epsfig{figure=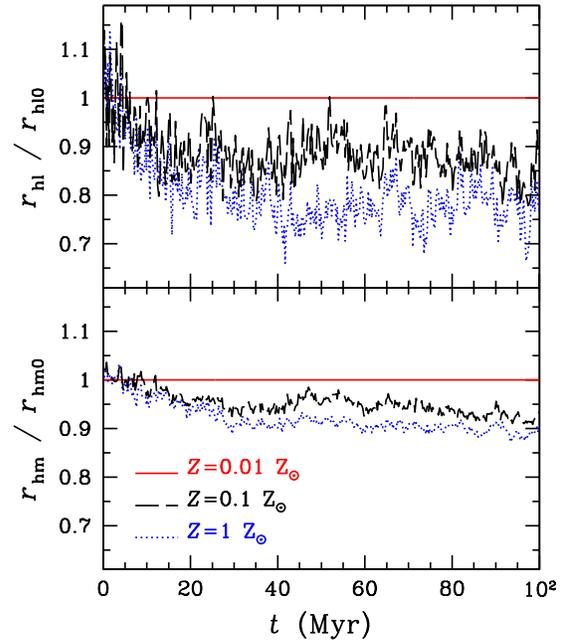,width=7.5cm} %was  fig4
}}
\caption{\label{fig:fig6}
Top panel: half-light radius ($r_{\rm hl}$) as a function of time for the three considered metallicities: $Z=0.01\,{}$Z$_\odot{}$ (solid red line,  A1), $Z=0.1\,{}$Z$_\odot{}$ (dashed black line,  B) and $Z=1\,{}$Z$_\odot{}$ (dotted blue line,  C). Each line in this Figure is the median value of 100 simulated SCs, and is normalized to the median half-light radius of SCs with $Z=0.01\,{}$Z$_\odot{}$ ($r_{\rm hl0}$). Bottom panel: half-mass radius ($r_{\rm hm}$) as a function of time for the three considered metallicities: $Z=0.01\,{}$Z$_\odot{}$ (solid red line,  A), $Z=0.1\,{}$Z$_\odot{}$ (dashed black line,  B) and $Z=1\,{}$Z$_\odot{}$ (dotted blue line,  C). Each line in this Figure is the median value of 100 simulated SCs, and is normalized to the median half-mass radius of SCs with $Z=0.01\,{}$Z$_\odot{}$ ($r_{\rm hm0}$).
}
\end{figure}
%%%%%%%%%%%%%%%%%%%%%%%%%%%%%%%%%%%%%%%%%%%%%%%%%%%%%%%%%%%%%%%%%%%%%%%%%%%%%%%
%%%%%%%%%%%%%%%%%%%%%%%%%%%%%%%%%%% FIGURE 7 %%%%%%%%%%%%%%%%%%%%%%%%%%%%%%%%%%
\begin{figure}
\center{{
\epsfig{figure=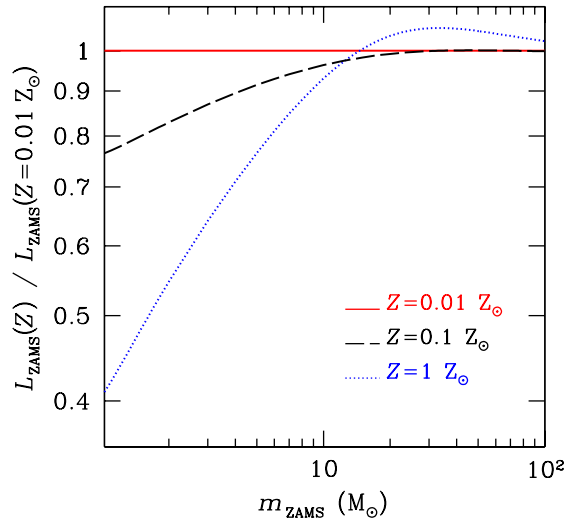,width=7.5cm} %was LTAMS.eps fig4bis
}}
\caption{\label{fig:fig7}
ZAMS luminosity ($L_{\rm ZAMS}$) versus ZAMS mass for $Z=0.01\,{}$Z$_\odot{}$ (solid red line), $Z=0.1\,{}$Z$_\odot{}$ (dashed black line) and $Z=1\,{}$Z$_\odot{}$ (dotted blue line). $L_{\rm ZAMS}$ is normalized to $L_{\rm ZAMS}(Z=0.01\,{}{\rm Z}_\odot{})$, i.e. to the ZAMS luminosity for $Z=0.01\,{}$Z$_\odot{}$.
}
\end{figure}
%%%%%%%%%%%%%%%%%%%%%%%%%%%%%%%%%%%%%%%%%%%%%%%%%%%%%%%%%%%%%%%%%%%%%%%%%%%%%%%
Fig.~\ref{fig:fig6} shows that the difference among  the half-light radii at different metallicity is larger ($\approx{}20$ per cent, although with larger fluctuations) than the difference among the half-mass radii. 
%Thus, our simulations indicate that young dense SCs with different metallicity differ for both the half-light radius and the half-mass radius.
 This results from the combination between mass segregation, and the metallicity dependence of the adopted stellar luminosity function. In our simulated SCs, mass segregation is very efficient, as shown by the fact that the core collapse occurs on a timescale shorter than the half-mass relaxation timescale (i.e., it is driven by dynamical friction, e.g. Portegies Zwart 2004). Thus, the region inside  the core radius is dominated by massive stars and remnants, while most low-mass stars are in the outer regions.

According to the fitting formulae by Hurley et al. (2000), a solar metallicity MS star with mass $\lesssim{}15$ M$_\odot{}$ is fainter than a sub-solar metallicity  MS star with the same mass. This is apparent from Fig.~\ref{fig:fig7}, where we compare the ZAMS luminosity  ($L_{\rm ZAMS}$) for stars with different metallicity. A similar difference persists during the entire MS. 
 Because of this difference in the luminosity function, and because our simulated SCs are mass-segregated, the light distribution tends to be more concentrated in metal-rich SCs than in metal-poor SCs.

\subsection{Other effects related to stellar evolution}
%%%%%%%%%%%%%%%%%%%%%%%%%%%%%%%%%%% FIGURE 8 %%%%%%%%%%%%%%%%%%%%%%%%%%%%%%%%%%
\begin{figure}
\center{{
\epsfig{figure=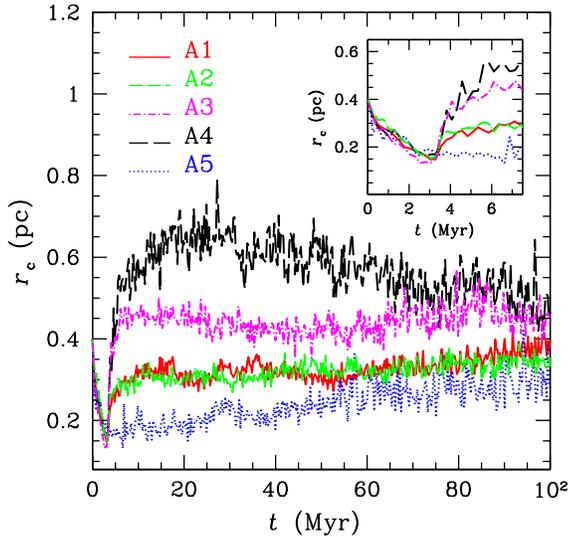,height=7.5cm} %was fig5
}}
\caption{\label{fig:fig8}
Core radius ($r_{\rm c}$) as a function of time for $Z=0.01\,{}$Z$_\odot{}$.  Solid red line:  A1, fiducial model with $f_{\rm PB}=0.1$ (the same as in Fig.~\ref{fig:fig3}); long dash-dotted green line:  A2, the same as the fiducial model, but with  $f_{\rm PB}=0.0$; short dash-dotted magenta line:  A3, $m_{\rm BH,\,{}max}=25$ M$_\odot{}$; dashed black line:  A4, BHs ejected by natal kick;  dotted blue line:  A5, no mass-loss by stellar winds. Each line in this Figure is the median value of a number of simulated SCs (see Table~2). In the insert: zoom of the first 7.5 Myr.
}
\end{figure}
%%%%%%%%%%%%%%%%%%%%%%%%%%%%%%%%%%%%%%%%%%%%%%%%%%%%%%%%%%%%%%%%%%%%%%%%%%%%%%%
%%%%%%%%%%%%%%%%%%%%%%%%%%%%%%%%%%% FIGURE 9 %%%%%%%%%%%%%%%%%%%%%%%%%%%%%%%%%%
\begin{figure}
\center{{
\epsfig{figure=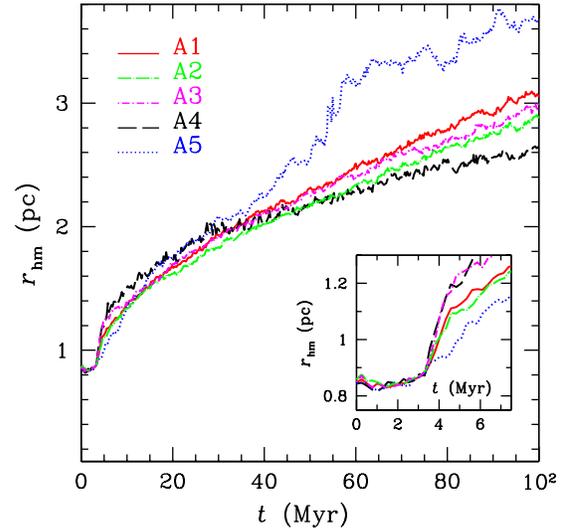,height=7.5cm}  %was fig6
}}
\caption{\label{fig:fig9}
Half-mass radius ($r_{\rm hm}$) as a function of time for $Z=0.01\,{}$Z$_\odot{}$. Solid red line:  A1 (the same as Fig.~\ref{fig:fig4}); long dash-dotted green line:  A2; short dash-dotted magenta line:  A3;  dashed black line:  A4; dotted blue line:  A5. Each line in this Figure is the median value of a number of simulated SCs (see Table~2). In the insert: zoom of the first 7.5 Myr.
}
\end{figure}
%%%%%%%%%%%%%%%%%%%%%%%%%%%%%%%%%%%%%%%%%%%%%%%%%%%%%%%%%%%%%%%%%%%%%%%%%%%%%%%

In this section, we estimate the impact on core collapse and post-core collapse phases of other effects connected with stellar evolution, including stellar remnants. To maximize the contribution of BHs to the reverse of core collapse, we consider SCs with $Z=0.01\,{}$Z$_\odot{}$, where the BH mass can be as high as $\sim{}80$ M$_\odot$.

The comparison between runs A1 (with $f_{\rm PB}=0.1$) and A2 (with $f_{\rm PB}=0$) shows that the existence of primordial binaries has almost no effect on both the core radius (Fig.~\ref{fig:fig8}) and the half-mass radius (Fig.~\ref{fig:fig9}). In fact, the very hard binaries needed to reverse the core collapse form by 3-body capture during the collapse phase even in the  $f_{\rm PB}=0$ runs.

The difference between runs A1 and A3 is the maximum mass of BHs: 80 and 25 M$_\odot{}$ for runs A1 and A3, respectively. This difference was achieved by ejecting impulsively  a large fraction of the stellar mass at the time of BH formation in runs A3. Thus, the core radius in runs A3 increases much faster than that in runs A1 because of this sudden mass-loss (Fig~\ref{fig:fig8}). On the other hand, the half-mass radius  (Fig~\ref{fig:fig9}) increases initially faster in runs A3 than in runs A1, but then slows down, because of the lower core density and because of the absence of massive BHs. In fact, massive BH binaries have a larger cross-section for three-body encounters, have (on average) a larger reservoir of internal energy and, thus, are more efficient in ejecting stars after three-body encounters (e.g. Downing 2012).
 
A similar but stronger trend can be observed by comparing runs A1 and A4. In runs A4 all BHs are removed from the simulation at birth, by assigning to them a natal kick of $10^3$ km s$^{-1}$. Immediately after core collapse, the core radius (Fig~\ref{fig:fig8}) and the half-mass radius (Fig~\ref{fig:fig9}) in runs A4 become significantly larger than in runs A1. In fact, the mass of all the BHs is completely lost from the potential well of the SC. On the other hand, the removal of all the BHs implies that less energy can be exchanged via three-body encounters. For this reason, the increase of the half-mass radius slows down at $t\gtrsim{}40$ Myr.

Finally, runs A5 represent the most extreme case, where mass-loss by stellar winds is switched off and the maximum BH mass is $110$ M$_\odot{}$ (the only difference between the BH mass and the ZAMS mass of the progenitor star comes from the recipes for the direct collapse of the star into a BH, see Section~\ref{sec:Method}). If mass-loss is switched off, the core collapse is much more dramatic: it lasts for $\gtrsim{}20$ Myr before that three-body encounters can reverse it (Fig~\ref{fig:fig8}). %Note that only in runs A5 gravothermal oscillations are well visible even after the reverse of core collapse, indicating that a stable equilibrium is much more difficult to achieve (e.g. Hut 1997). 
The half-mass radius (Fig~\ref{fig:fig9}) expands dramatically after core collapse, because of the amount of energy pumped into the halo by three-body encounters.

\section{Conclusions}\label{sec:conclu}
In this paper, we discussed the effects of metallicity on the core collapse and post-core collapse phase in young intermediate-mass SCs. This was done by means of $N$-body simulations including recipes for metallicity-dependent stellar evolution, stellar winds and formation of stellar remnants.

In the simulated SCs, core collapse is almost coincident with the peak of mass-loss by stellar winds. Metal-rich SCs lose more mass by stellar winds than metal-poor SCs: the reversal of core collapse is faster and stronger in the former with respect to the latter. On the other hand, the difference in the core radius among metal-rich and metal-poor SCs decreases as soon as mass losses by massive stars are over. In the later stages of SC life, the core evolution is ruled mainly by three-body encounters.

Since the reversal of core collapse is slower in metal-poor SCs, the half-mass radius expands more in metal-poor SCs than in metal-rich SCs. The maximum difference among the  half-mass radii of $Z=0.01$ and $Z=1$ Z$_\odot$ SCs is $\sim{}10$ per cent. When considering the half-light radii rather than the half-mass radii, the difference at late times is larger ($\approx{}20$ per cent). This is a consequence of the metallicity dependence of the adopted stellar luminosity function. %In agreement with Jord\'an (2004), this is due to the fact that metal-poor stars lose less mass and have slightly longer MS lifetimes than metal-rich stars with the same ZAMS mass.

We also checked the effects of other aspects of stellar evolution. When stellar winds are completely suppressed, the core collapse phase lasts much longer and the half-mass radius in the post-core collapse phase can be even 20 per cent larger than in the fiducial model for the same metallicity.
If all the BHs are ejected by natal kick, the core initially expands faster because of the impulsive mass-loss, but all the massive binaries are lost from the SC, quenching the effects of three-body encounters. 

In the last few years, young SCs were at the centre of important observational campaigns (e.g. Bica et al. 2003; Mercer et al. 2005; Borissova et al. 2011; Bianchi et al. 2012; Richards et al. 2012;  see Portegies Zwart et al. 2010 for a recent review), and are one of the main targets of the ongoing Gaia ESO survey (Gilmore et al. 2012). Thus, comparing the predictions of our simulations with the current and forthcoming data about metallicity and half-light radii of young SCs will give new insights on the formation and dynamical evolution of SCs. 

\section*{Acknowledgments}
We thank the referee, Peter Anders, for his comments that improved the manuscript. We made use of the public software package {\sc starlab} (version 4.4.4) and of the SAPPORO library (Gaburov, Harfst \&{} Portegies Zwart 2009) to run {\sc starlab} on graphics processing units (GPUs). We acknowledge all the developers of {\sc starlab}, and especially its primary authors: Piet Hut, Steve McMillan, Jun Makino, and Simon Portegies Zwart. We thank the authors of SAPPORO, and in particular E. Gaburov, S. Harfst and S.  Portegies Zwart. We also thank Emanuele Ripamonti, Luca Zampieri, Marica Branchesi, Paola Marigo, Alessia Gualandris, Antonella Vallenari and Rosanna Sordo for stimulating discussions.
 We acknowledge the CINECA Award N.  HP10B3BJEW, HP10CLI3BX, HP10CXB7O8 and HP10C894X7, 2011 for the availability of high performance computing resources and support. MM acknowledges financial support from INAF through grant PRIN-2011-1. %LZ acknowledges financial support from ASI/INAF grant no. I/009/10/0. 
AB acknowledges financial support from MIUR through grant PRIN-2009-1. MM thanks the Aspen Center for Physics, where part of this work was done.

%\appendix
%\section{The importance of primordial binaries}

\begin{thebibliography}{}
\bibitem{}Aarseth S. J., 2012, MNRAS, 422, 841
%\bibitem{}Akiyama Sh., Wheeler J. C., Meier D. L., Lichtenstadt I., 2003, ApJ, 584, 954
\bibitem{}Anders P., Baumgardt H., Bissantz N., Portegies Zwart S., 2009, MNRAS, 395, 2304
\bibitem{}Angeletti L., Giannone P., 1977, A\&{}A, 58, 363
\bibitem{}Angeletti L., Giannone P., 1980, A\&{}A, 85, 113
\bibitem{}Applegate J. H., 1986, ApJ, 301, 132 
%%\bibitem{}Appleton P. N., Marston A. P., 1997, AJ, 113, 201
%%%\bibitem{}Anders E., Grevesse N., 1989, Geochimica et Cosmochimica Acta 53, 197
%\bibitem{}Ardeljan N. V., Bisnovatyi-Kogan G. S., Moiseenko S. G., 2005, MNRAS, 359, 333
\bibitem{}Baraffe I., Heger A., Woosley S. E., 2001, ApJ, 550, 890
\bibitem{}Barmby P., Holland S., Huchra J. P., 2002, AJ, 123, 1937
%\bibitem{}Begelman M. C., 2002, ApJ, 568L, 97
%%\bibitem{}Belczynski K., Kalogera V., Bulik T., 2002, ApJ, 572, 407
%\bibitem{}Belczynski K., Sadowski A., Rasio F. A., 2004, ApJ, 611, 1068
%\bibitem{}Belczynski K., Kalogera V., Zezas A., Fabbiano G., 2004, ApJ, 601, L147 
%%\bibitem{}Belczynski K., Taam R. E., Kalogera V., Rasio F. A., Bulik T., 2007, ApJ, 662, 504
%\bibitem{}Belczynski K., Kalogera V., Rasio F. A., Taam R. E., Zezas A., Bulik T., Maccarone Th. J., Ivanova N., 2008, ApJS, 174, 223
\bibitem{}Belczynski K., Bulik T., Fryer C. L., Ruiter A., Valsecchi F., Vink J. S., Hurley J. R., 2010, ApJ, 714, 1217 (B10)
\bibitem{}Belkus H., Van Bever J., Vanbeveren D. 2007, ApJ, 659, 1576
%%\bibitem{}Bell E. F., 2003, ApJ, 586, 794
%%\bibitem{}Bell E. F., Kennicutt R. C. Jr., 2001, ApJ, 548, 681
%\bibitem{}Berghea C., 2009, {\it Ultraluminous X-ray sources and their Environment}, PhD thesis, {\tt http://students.cua.edu/79berghea/papers/dissertation.pdf}
%%\bibitem{}Bertola F., Bressan A., Burstein D., Buson L. M., Chiosi C., di Serego Alighieri S., 1995, ApJ, 438, 680
%%\bibitem{}Bevington P. R., 1969, {\it Data reduction and error analysis for the physical sciences}, New York: McGraw-Hill
\bibitem{}Bianchi L., Efremova B., Hodge P., Kang Y., 2012, AJ, 144, 142
\bibitem{}Bica E., Dutra C. M., Soares J., Barbuy B., 2003, A\&{}A, 404, 223
\bibitem{}Binney J., Tremaine S., 1987, Galactic Dynamics. Princeton Univ. Press, Princeton , NJ 
%%%\bibitem{}Bird A. J., et al., 2006, ApJ, 636, 765
%\bibitem{}Blecha L., Ivanova N., Kalogera V., Belczynski K., Fregeau J., Rasio F., 2006, ApJ, 642, 427 (B06)
%\bibitem{}B\"oker T., van der Marel R. P., Vacca W. D., 1999, AJ, 118, 831
%\bibitem{}Bonnell \&{} Kroupa, 1998, to appear in The Orion Complex Revisited, eds. M.J.McCaughrean, A.Burkert, ASP Conference Series, arXiv:astro-ph/9802306
\bibitem{}Borissova J., et al., 2011, A\&{}A, 532A, 131
%\bibitem{}Brandt N., Podsiadlowski Ph., 1995, MNRAS, 274, 461
%\bibitem{}Brandt W. N., Podsiadlowski Ph., Sigurdsson S., 1995, MNRAS, 277, L35
%%\bibitem{}Brassington N. J., Read A. M., Ponman T. J., 2005, MNRAS, 360, 801 
%%\bibitem{}Brassington N. J., Ponman T. J., Read A. M., 2007, MNRAS, 377, 1439 
%\bibitem{}Bresolin F., Kudritzki R. P., 2004, in  Origin and Evolution of the Elements, from the Carnegie Observatories Centennial Symposia. Published by Cambridge University Press, as part of the Carnegie Observatories Astrophysics Series. Edited by McWilliam A.  and Rauch M., 283, arXiv:astro-ph/0303620
%%\bibitem{}Bressan A., Fagotto F., Bertelli G., Chiosi C., 1993, A\&{}AS, 100, 647
\bibitem{}Bromm V., Kudritzki R. P., Loeb A., 2001, ApJ, 552, 464
%%\bibitem{}Buat V., Boselli A., Gavazzi G., Bonfanti C., 2002, A\&{}A, 383, 801
%\bibitem{}Carpano S., Pollock A. M. T., Prestwich A., Crowther P., Wilms J., Yungelson L., Ehle M., 2007, A\&{}A, 466, L17
%%\bibitem{}Cash W., 1979, ApJ, 228, 939
%%\bibitem{}Castro-Tirado A. J., Brandt S., Lund N., 1992, IAU Circ., 5590, 2. Edited by Green D. W. E.
%\bibitem{}Charles P. A., Coe M. J., 2006, In: Compact stellar X-ray sources. Edited by Walter Lewin \&{} Michiel van der Klis. Cambridge Astrophysics Series, No. 39. Cambridge, UK: Cambridge University Press, 215
%%\bibitem{}Chatterjee S., Goswami S., Umbreit S., Glebbeek E., Rasio F. A., Hurley J., 2009, ApJ, submitted, arXiv:0911.1483
%\bibitem{}Chatterjee S., Fregeau J. M., Umbreit S., Rasio F. A., 2010, ApJ, 719, 915
\bibitem{}Chernoff D. F., Shapiro S. L., 1987, ApJ, 322, 113
\bibitem{}Chernoff D. F., Weinberg M. D., 1990, ApJ, 351, 121
%%\bibitem{}Clemens M. S., Alexander P., Green D. A., 1999, MNRAS, 307, 481 
%%\bibitem{}Colbert E. J. M., Petre R., Schlegel E. M., Ryder S. D., 1995, ApJ, 446, 177
%%\bibitem{}Colbert E. J. M., Heckman T. M., Ptak A. F., Strickland D. K., Weaver K. A., 2004, ApJ, 602, 231
%%%\bibitem{}Colbert E. J. M., Miller M. C., 2005, Invited review talk at the Tenth Marcel Grossmann Meeting on General Relativity, Rio de Janeiro, July 20-26, 2003. Proceedings edited by M. Novello, S. Perez-Bergliaffa and R. Ruffini, World Scientific, Singapore, 2005
%\bibitem{}Colpi M., Mapelli M., Possenti A., 2003, ApJ, 599, 1260
%%\bibitem{}Condon J. J., Cotton W. D., Broderick J. J., 2002, AJ, 124, 675
%\bibitem{}Copperwheat C., Cropper M., Soria R., Wu K., 2007, MNRAS, 376, 1407
%%\bibitem{}Corwin H. G. Jr., Buta R. J., de Vaucouleurs G., 1994, AJ, 108, 2128
%%\bibitem{}Cropper M., Soria R., Mushotzky R. F., Wu K., Markwardt C. B., Pakull M., 2004, MNRAS, 349, 3
%\bibitem{}Crowther P. A., Carpano S., Hadfield L. J., Pollock A. M. T., 2007, A\&{}A, 469, L31
%\bibitem{}Crowther P. A., Barnard R., Carpano S., Clark J. S., Dhillon V. S., Pollock A. M. T., 2010, MNRAS, 403, L41 
%%\bibitem{}D'Agostini G., 2005, arXiv:physics/0511182
%%%\bibitem{}Daflon S., Cunha K., 2004, ApJ, 617, 1115
%\bibitem{}Davies M. B., Benz W., Hills J. G., 1994, ApJ, 424, 870
%\bibitem{}Davies M. B., 1995, MNRAS, 276, 887
%\bibitem{}Davies M. B., 2002, in `Omega Centauri, A Unique Window into Astrophysics'. ASP Conference Proceedings, Vol. 265. Edited by Floor van Leeuwen, Joanne D. Hughes, and Giampaolo Piotto. San Francisco, Astronomical Society of the Pacific, p.215
%%%\bibitem{}Deegan P., Combet C., Wynn G. A., 2009, MNRAS, in press, arXiv:0908.2566
%\bibitem{}Delgado-Donate E. J., Clarke C. J., Bate M. R., Hodgkin S. T., 2004, MNRAS, 351, 617
%\bibitem{}Dessart L., O'Connor E., Ott Ch., 2012, ApJ, 754, 76
%%%\bibitem{}de Vaucouleurs G., de Vaucouleurs A., Corwin H. G. Jr., Buta R. J., Paturel G., Fouque P., 1991, {\it Third Reference Catalogue of Bright Galaxies}, Springer-Verlag Berlin Heidelberg New York 
%\bibitem{}D'Ercole A., D'Antona F., Carini R., Vesperini E., Ventura P., 2012, MNRAS, 423, 1521
%%%\bibitem{}Dewangan G. C., Miyaji T., Griffiths R. E., Lehmann I., 2004, ApJ, 608L, 57
\bibitem{}Dias W. S., Alessi B. S., Moitinho A., L\'epine J. R. D., 2002, A\&{}A, 389, 871
%%%\bibitem{}Dickey J. M., Lockman F. J., 1990, ARA\&{}A, 28, 215 [DL90]
%%%\bibitem{}Dopita M. A., Ryder S. D., 1994, ApJ, 430, 163
%%%\bibitem{}Dors O. L. Jr., Copetti M. V. F., 2005, A\&{}A, 437, 837
\bibitem{}Downing J. M. B., Benacquista M. J., Giersz M., Spurzem R., 2010, MNRAS, 407, 1946
%\bibitem{}Downing J. M. B., Benacquista M. J., Giersz M., Spurzem R., 2011, MNRAS, 416, 133
\bibitem{}Downing J. M. B., 2012, MNRAS, 425, 2234
%\bibitem{}Dray L. M., 2006, MNRAS, 370, 2079
\bibitem{}Duquennoy A., \&{} Mayor M., 1991, A\&{}A, 248, 485
%%%\bibitem{}Edmunds M. G., Pagel B. E. J., 1984, MNRAS, 211, 507
%%%\bibitem{}Eldridge J. J., Tout C. A., 2004, MNRAS, 353, 87
%%%\bibitem{}Eldridge J. J., Vink J. S., 2006, A\&{}A, 452, 295
\bibitem{}Elson R., Hut P., Inagaki Sh., 1987, ARA\&{}A, 25, 565
%\bibitem{}Evans C. J., et al., 2005, A\&{}A, 437, 467
%\bibitem{}Evans C. J., Lennon D. J., Smartt S. J., Trundle C., 2006, A\&{}A, 456, 623 
%%%\bibitem{}Fabbiano G., Zezas A., Murray S. S., 2001, ApJ, 554, 1035
%%%%\bibitem{}Fabbiano G., et al., 2004, ApJ, 605L, 21
%%%\bibitem{}Fabbiano G., Baldi A., King A. R., Ponman T. J., Raymond J., Read A., Rots A., Schweizer F., Zezas A., 2004, ApJ, 605L, 21
%%%\bibitem{}Fagotto F., Bressan A., Bertelli G., Chiosi C., 1994a, A\&{}AS, 105, 29
%%%\bibitem{}Fagotto F., Bressan A., Bertelli G., Chiosi C., 1994b, A\&{}AS, 105, 39
%%%\bibitem{}Fender R., Belloni T., 2004, ARA\&{}A, 42, 317
%%%\bibitem{}Fosbury R. A. E., Hawarden T. G., 1977, MNRAS, 178, 473
%%\bibitem{}Fragos T., Willems B., Kalogera V., Ivanova N., Rockefeller G., Fryer C. L., Young P. A., 2009, ApJ, 697, 1057
%\bibitem{}Fregeau J. M., G\"urkan M. A., Joshi K. J., Rasio F. A., 2003, ApJ, 593, 772
%\bibitem{}Fregeau J. M., \&{} Rasio F. A., 2007, ApJ, 658, 1047
%%\bibitem{}Freitag M., G\"urkan M. A., Rasio F. A., 2006, MNRAS, 368, 141
%%%\bibitem{}French, H. B., 1980, ApJ, 240, 41
%%%%\bibitem{}Fridriksson J. K., Homan J., Lewin W. H. G., Kong A. K. H., Pooley D., 2008, ApJS, 177, 465
\bibitem{}Fryer Ch. L., 1999, ApJ, 522, 413
%\bibitem{}Fryer Ch. L., \&{} Heger A., 2000, ApJ, 541, 1033
\bibitem{}Fryer Ch. L., Kalogera V., 2001, ApJ, 554, 548
%\bibitem{}Fryer Ch. L., \&{}  Warren M. S., 2004, ApJ, 601, 391 
\bibitem{}Fryer Ch. L., Belczynski K., Wiktorowicz G., Dominik M., Kalogera V., Holz D. E., 2012, ApJ, 749, 91
\bibitem{}Gaburov E., Harfst S., Portegies Zwart S., 2009, New Astronomy, 14, 630
%%\bibitem{}Gaburov E., Lombardi J. C. Jr., Portegies Zwart S., 2010, MNRAS, 402, 105
%%%%\bibitem{}Gao Y., Wang Q. D., Appleton P. N., Lucas R. A., 2003, ApJ, 596L, 171
%\bibitem{}Garnett D. R., 1990, ApJ, 363, 142
%%%\bibitem{}Gavazzi G., Boselli A., Pedotti P., Gallazzi A., Carrasco L., 2002, A\&{}A, 396, 449
%%%\bibitem{}Gehrels N., 1986, ApJ, 303, 336
%%\bibitem{}Gelatt A. E., Hunter D. A., Gallagher J. S., 2001, The Publications of the Astronomical Society of the Pacific, Volume 113, Issue 780, p. 142
%%%%\bibitem{}Ghosh K. K., Mapelli M., 2008, MNRAS, 386L, 38
\bibitem{}Gieles M., Portegies Zwart S. F., 2011, MNRAS, 410, L6
%\bibitem{}Giersz M., 1998, MNRAS, 298, 1239 
%%%\bibitem{}Gil de Paz A., Madore B. F., Pevunova O., 2003, ApJS, 147, 29
%%%\bibitem{}Gilfanov M., Grimm H.-J., Sunyaev R., 2004a, MNRAS, 347L, 57
%%%\bibitem{}Gilfanov M., Grimm H.-J., Sunyaev R., 2004b, Nuclear Physics B Proceedings Supplements, 132, 369  
%%%\bibitem{}Gilfanov M., Grimm H.-J., Sunyaev R., 2004c, MNRAS, 351, 1365
\bibitem{}Gilmore G., 2012, The Messenger, 147, 25
%%\bibitem{}Gon\c{c}alves A. C., Soria R., 2006, MNRAS, 371, 673
%%%\bibitem{}Gonzalez-Delgado R. M., Perez E., Diaz A. I., Garcia-Vargas M. L., Terlevich E., Vilchez J. M., 1995, ApJ, 439, 604
%%%%%\bibitem{}Grevesse N., Asplund M., Sauval A. J., 2007, Space Sci. Rev., 130, 105
%%%\bibitem{}Grimm H.-J., Gilfanov M., Sunyaev R., 2002, A\&{}A 391, 923
%%%\bibitem{}Grimm H.-J., Gilfanov M., Sunyaev R., 2003, MNRAS, 339, 793
%%\bibitem{}Gris\'e F., Pakull M. W., Motch C., 2006, in Populations of High Energy Sources in Galaxies, Edited by Meurs E. J. A., Fabbiano G., Cambridge: Cambridge University Press, p.302
%\bibitem{}Gris\'e F., Pakull M. W., Soria R., Motch C., Smith I. A., Ryder S. D., B\"ottcher M., 2008, A\&{}A, 486, 151
%\bibitem{}Gris\'e F., Kaaret P., Pakull M. W., Motch C., 2011, ApJ, 734, 23
%%\bibitem{}Gualandris A., Colpi M., Portegies Zwart S., Possenti A., 2005, ApJ, 618, 845
%\bibitem{}G\"urkan M. A., Freitag M., Rasio F. A., 2004, ApJ, 604, 632
%%%%\bibitem{}G\"urkan M. A., Fregeau J. M., Rasio F. A., 2006, ApJ, 640L, 39
%%%\bibitem{}Guseva N. G., Izotov Y. I., Thuan T. X., 2000, ApJ, 531, 776
%%%%\bibitem{}Gvaramadze V. V., Gualandris A., Portegies Zwart S., 2008, MNRAS, 385, 929
\bibitem{}Hamann W.-R., Koesterke L., 1998, A\&{}A, 335, 1003
%\bibitem{}Harlaftis E. T., Greiner J., 2004, A\&{}A, 414L, 13
\bibitem{}Harris W. E., Harris G. L. H., Holland S. T., McLaughlin D. E., 2002, AJ, 124, 1435
\bibitem{}Harris W. E., 2009, ApJ, 703, 939
%%%\bibitem{}Hasinger G., Burg R., Giacconi R., Schmidt M., Tr\"umper J., Zamorani G., 1998, A\&{}A, 329, 482
%%%\bibitem{}Hawley S. A., Phillips M. M., 1980, ApJ, 235, 783
\bibitem{}Heger A., Fryer C.L., Woosley S.E., Langer N., Hartmann D.H., 2003a, ApJ, 591, 288
%\bibitem{}Heger A., Woosley S.E., Fryer C.L., Langer N., 2003b, in From Twilight to Highlight: The Physics of Supernovae: Proceedings of the ESO/MPA/MPE Workshop Held at Garching, Germany, 29-31 Juli 2002, ESO ASTROPHYSICS SYMPOSIA. ISBN 3-540-00483-1. Edited by W. Hillebrandt and B. Leibundgut. Springer-Verlag, p. 3
%%\bibitem{}Heger A., Woosley S.E., 2002, ApJ, 567, 532 
\bibitem{}Heggie D. C., 1975, MNRAS, 173, 729
%\bibitem{}Heggie D. C., Hut P., 1993, ApJS, 85, 347
%%\bibitem{}Heggie D. C., Rasio F. A., 1996, MNRAS, 282, 1064
%%%\bibitem{}Helou G., et al., 2004, ApJS, 154, 253
%%%\bibitem{}Helmboldt J. F., Walterbos R. A. M., Bothun G. D., O'Neil K., de Blok W. J. G., 2004, ApJ, 613, 914 
%\bibitem{}H\'enon M. H., 1971a, Ap\&{}SS, 13, 284
%\bibitem{}H\'enon M. H., 1971b, Ap\&{}SS, 14, 151
%%%%\bibitem{}Hernquist L., Weil M. L., 1993, MNRAS, 261, 804
%%%%\bibitem{}Higdon J. L., 1995, ApJ, 455, 524
%%%%\bibitem{}Higdon J. L., 1996, ApJ, 467, 241
%%%%\bibitem{}Higdon J. L., Wallin J. F., 1997, ApJ, 474, 686
\bibitem{}Hillenbrand L. A., Hartmann L. W., 1998, ApJ, 492, 540
%\bibitem{}Hills J. G., Fullerton L. W., 1980, AJ, 85, 1281
%%%\bibitem{}Hirashita H., Buat V., Inoue A. K., 2003, A\&{}A, 410, 83
%%%\bibitem{}Hopkins A. M., Schulte-Ladbeck R. E., Drozdovsky I. O., 2002, AJ, 124, 862
%%%%\bibitem{}Horellou C., Combes F., 2001, Ap\&{}SS, 276, 1141
\bibitem{}Humphreys R. M., Davidson K., 1994, Astronomical Society of the Pacific, Publications, 106, 1025
%%%\bibitem{}Hunter D. A., Elmegreen B. G., 2004, AJ, 128, 2170
%%%%\bibitem{}Hunter D. A., Gillett F. C., Gallagher J. S. III, Rice W. L., Low F. J., 1986, ApJ, 303, 171
\bibitem{}Hurley J. R., Pols O. R., Tout C. A., 2000, MNRAS, 315, 543
\bibitem{}Hurley J. R., Tout C. A., Aarseth S. J., Pols O. R., 2004, MNRAS, 355, 1207
\bibitem{}Hurley J. R., 2007, MNRAS, 379, 93 
\bibitem{}Hurley J. R., Shara M. M., 2012, MNRAS, 425, 2872
%\bibitem{}Hurley J. R., Tout C. A., Pols O. R., 2002, MNRAS, 329, 897
%\bibitem{}Hut P., 1997, Gravitational Thermodynamics, arXiv:astro-ph/9704286
%\bibitem{}Hut P., McMillan S., Goodman J., Mateo M., Phinney E. S., Pryor C., Richer H. B., Verbunt F., Weinberg M., 1992, Astronomical Society of the Pacific, Publications, 104, 981
%%%\bibitem{}Iglesias-P\'aramo J., et al., 2006, ApJS, 164, 38
%%%\bibitem{}Irwin J. A., Sarazin C. L., Bregman J. N., 2002, ApJ, 570, 152
%%\bibitem{}Irwin J. A., Bregman J. N., Athey A. E., 2004, ApJ, 601L, 143
%\bibitem{}Ivanova N., Belczynski K., Fregeau J. M., Rasio F., 2005, MNRAS, 358, 572
%\bibitem{}Ivanova N., Heinke C. O., Rasio F. A., Taam R. E., Belczynski K., Fregeau J., 2006, MNRAS, 372, 1043
%\bibitem{}Ivanova N., Heinke C. O., Rasio F. A., Belczynski K., Fregeau J., 2008, MNRAS, 386, 553
%%%%\bibitem{}Izotov Y. I., Stasi\'nska G., Meynet G., Guseva N. G., Thuan T. X., 2006, A\&{}A, 448, 955
%%%\bibitem{}Izotov Y. I., Thuan T. X., Lipovetsky V. A., 1997, ApJS, 108, 1
\bibitem{}Jord\'an A., 2004, ApJ, 613, L117
\bibitem{}Jord\'an A., et al., 2005, ApJ, 634, 1002
\bibitem{}Jord\'an A., et al., 2009, ApJS, 180, 54
%\bibitem{}Joshi K. J., Rasio F. A., Portegies Zwart S., 2000, ApJ, 540, 969
%\bibitem{}Joshi K. J., Nave C. P., Rasio F. A., 2001, ApJ, 550, 691 
%%%\bibitem{}Kaaret P., 2001, ApJ, 560, 715
%%%\bibitem{}Kaaret P., Alonso-Herrero A., 2008, ApJ, 682, 1020
%%\bibitem{}Kaaret P., Ward M. J., Zezas A., 2004a, MNRAS, 351L, 83
%\bibitem{}Kaaret P., Alonso-Herrero A., Gallagher J. S., Fabbiano G., Zezas A., Rieke M. J., 2004, MNRAS, 348L, 28
%%%\bibitem{}Kalogera V., Belczynski K., Kim C., O'Shaughnessy R., Willems B., 2007, PhR, 442, 75
%%%\bibitem{}Kehrig C., Telles E., Cuisinier F., 2004, AJ, 128, 1141
%%%\bibitem{}Kennicutt R. C. Jr., Kent, S. M., 1983, AJ, 88, 1094
%%%\bibitem{}Kennicutt R. C. Jr., Roettiger K. A., Keel W. C., van der Hulst J. M., Hummel E., 1987, AJ, 93, 1011
%%%\bibitem{}Kennicutt R. C. Jr., 1992, ApJ, 388, 310
%%%\bibitem{}Kennicutt R. C. Jr., Tamblyn P., Congdon C. E., 1994, ApJ, 435, 22
%%%\bibitem{}Kennicutt R. C. Jr., 1998, ApJ, 498, 541 [K98]
%%\bibitem{}Kennicutt R. C. Jr., Bresolin F., Garnett D. R., 2003, ApJ, 591, 801
%%%\bibitem{}Kewley L. J., Geller M. J., Jansen R. A., Dopita M. A., 2002, AJ, 124, 3135
%%%%\bibitem{}Kewley L. J., Geller M. J., Jansen R. A., 2004, AJ, 127, 2002
%%%\bibitem{}Kewley L. J., Jansen R. A., Geller M. J., 2005, PASP, 117, 227
\bibitem{}King I. R., 1966, AJ, 71, 64
%%\bibitem{}King A. R., Davies M. B., Ward M. J., Fabbiano G., Elvis M., 2001, ApJL, 552, 109 
%%\bibitem{}King A. R., Pounds K. A., 2003, MNRAS, 345, 657
%%%%\bibitem{}King A. R., 2004, MNRAS, 347L, 18
%%\bibitem{}King A. R., 2008, MNRAS, 385L, 113
%%\bibitem{}King I. R., 1966, AJ, 71, 64
\bibitem{}Kraicheva Z. T., Popova E. I., Tutukov A. V., Iungelson L. R., 1978, Astronomicheskii Zhurnal, 55, 1176
\bibitem{}Kroupa P., 2001, MNRAS, 322, 231
\bibitem{}Kudritzki R. P., Pauldrach A., Puls J., 1987, A\&{}A, 173, 293
\bibitem{}Kudritzki R.-P., Puls J., 2000, ARA\&{}A, 38, 613
\bibitem{}Kudritzki R. P., 2002, ApJ, 577, 389
\bibitem{}Kuhn M. A., Baddeley A., Feigelson E. D., Getman K. V., Broos P. S., Townsley L. K., Povich M. S., Naylor T., King R. R., Busk H. A., Luhman K. L., 2012, arXiv:1208.3492, to appear in The Labyrinth of Star Formation, (eds.) D. Stamatellos, S. Goodwin, and D. Ward-Thompson, Springer, in press
%%\bibitem{}Kulkarni S. R., Hut P., McMillan S.,  1993, Nature, 364, 421
\bibitem{}Kundu A., Whitmore B. C., 1998, AJ, 116, 2841
\bibitem{}Kundu A., Whitmore B. C., 2001, AJ, 121, 2950
\bibitem{}Kundu A., Whitmore B. C., Sparks W. B., Macchetto F. D., Zepf S. E., Ashman K. M., 1999, ApJ, 513, 733
%\bibitem{}Lada Ch. J., Lada E. A., 2003, ARA\&{}A, 41, 57
\bibitem{}Larsen S. S., Brodie J. P., Huchra J. P., Forbes D. A., Grillmair C. J., 2001, AJ, 121, 2974
\bibitem{}Larsen S. S., Forbes D. A., Brodie J. P., 2001, MNRAS, 327, 1116
%\bibitem{}Lee C.-H., Brown G. E., Wijers R. A. M. J., 2002, ApJ, 575, 996
%%%\bibitem{}Lee H., Skillman E. D., 2004, ApJ, 614, 698
%%%\bibitem{}Legrand F., Tenorio-Tagle G., Silich S., Kunth D., Cervi$\tilde{\rm n}$o M., 2001, ApJ, 560, 630
\bibitem{}Leitherer C., Robert C., Drissen L., 1992, ApJ, 401, 596
%%%%%\bibitem{}Leitherer C., Li I.-H., Calzetti D., Heckman T. M., 2002, ApJS, 140, 303
%\bibitem{}Linden T., Kalogera V., Sepinsky J. F., Prestwich A., Zezas A., Gallagher J., 2010, ApJ, 725, 1984 (L10)
%%%%\bibitem{}L\'ipari S., D\'iaz R., Taniguchi Y., Terlevich R., Dottori H., Carranza G., 2000, AJ, 120, 645
%%%%\bibitem{}Lira P., Ward M., Zezas A., Alonso-Herrero A., Ueno S., 2002, MNRAS, 330, 259
%%%\bibitem{}Liu J.-F., Bregman J. N., 2005, ApJS, 157, 59 (LB05)
%%%\bibitem{}Liu Q. Z., Mirabel I. F., 2005, A\&{}A, 429, 1125 (LM05)
%%%\bibitem{}Liu J.-F., Bregman J. N., Irwin J., 2006, ApJ, 642, 171
%\bibitem{}Liu J.-F., Bregman J., Miller J., Kaaret Ph., 2007, ApJ, 661, 165
%\bibitem{}Livio M., Warner B., 1984, The Observatory, 104, 152
%%%\bibitem{}Long K.S., Helfand D.J., Grabelsky D.A., 1981, ApJ, 248, 925
%%%\bibitem{}Lonsdale C. J., Helou G., 1985, {\it Cataloged Galaxies and Quasars Observed in the IRAS Survey}, Pasadena: Jet Propulsion Laboratory (JPL)
%%%\bibitem{}L\'opez-Corredoira M., Guti\'errez C. M., 2006, A\&{}A, 454, 77
\bibitem{}Loup C., Forveille T., Omont A., Paul J. F., 1993, Astronomy and Astrophysics Supplement Series, 99, 291
%\bibitem{}Madhusudhan N., Justham S., Nelson L., Paxton B., Pfahl E., Podsiadlowski Ph., Rappaport S., 2006, ApJ, 640, 918
%\bibitem{}Madhusudhan N., Rappaport S., Podsiadlowski Ph., Nelson L., 2008, ApJ, 688, 1235
%\bibitem{}Maeder A., Meynet G., 2010, New Astronomy, 54, 32
%\bibitem{}Maeder A., Meynet G., 2012, Reviews of Modern Physics, 84, 25
\bibitem{}Maeder A., 1992, A\&{}A, 264, 105 (M92)
\bibitem{}Mapelli M., Sigurdsson S., Colpi M., Ferraro F. R., Possenti A., Rood R. T., Sills A., Beccari G., 2004, ApJ, 605L, 29
\bibitem{}Mapelli M., Sigurdsson S., Ferraro F. R., Colpi M., Possenti A., Lanzoni B., 2006, MNRAS, 373, 361
\bibitem{}Mapelli M., Colpi M., Zampieri L., 2009a, MNRAS, 395L, 71
\bibitem{}Mapelli M., Ripamonti E., Battaglia G., Tolstoy E., Irwin M. J., Moore B., Sigurdsson S., 2009b, MNRAS, 396, 1771
%\bibitem{}Mapelli M., Ripamonti E., Zampieri L., Colpi M., Bressan A., 2010, MNRAS, 408, 234
%\bibitem{}Mapelli M., Ripamonti E., Zampieri L., Colpi M., 2011a, AN, 332, 414
%\bibitem{}Mapelli M., Ripamonti E., Zampieri L., Colpi M., 2011b, MNRAS, 416, 1756
\bibitem{}Mapelli M., Zampieri L., Ripamonti E., Bressan A., 2013, MNRAS, in press, arXiv1211.6441M (M13)
%%%%\bibitem{}Mapelli M., Moore B., Giordano L., Mayer L., Colpi M., Ripamonti E., Callegari S., 2008a, MNRAS, 383, 230
%%%%\bibitem{}Mapelli M., Moore B., Ripamonti E., Mayer L., Colpi M., Giordano L., 2008b, MNRAS, 383, 1223
%\bibitem{}Marks M., Kroupa P., Dabringhausen J., Pawlowski M. S., 2012, MNRAS, 422, 2246
%%%%\bibitem{}Marston A. P., Appleton P. N., 1995, AJ, 109, 1002
%%%\bibitem{}Martin C. L., 1997, ApJ, 491, 561
%%%\bibitem{}Martin P., Friedli D., 1997, A\&{}A, 326, 449
\bibitem{}Martins F., Hillier D. J., Paumard T., Eisenhauer F., Ott T., Genzel R., 2008, A\&{}A, 478, 219
%%%\bibitem{}Masegosa J., Moles M., Campos-Aguilar A., 1994, ApJ, 420, 576
%%%%%\bibitem{}Matteucci F., Tosi M., 1985, MNRAS, 217, 391
%%%\bibitem{}Mayya Y. D., Bizyaev D., Romano R., Garcia-Barreto J. A., Vorobyov E. I., 2005, ApJ, 620L, 35
\bibitem{}Mercer E. P., et al., 2005, ApJ, 635, 560
%\bibitem{}Merritt D., 2001, ApJ, 556, 245
%%%%\bibitem{}Meynet G., Maeder A., 2003, A\&{}A, 404, 975
%%%%%\bibitem{}Mihos J. C., Bothun G. D., Richstone D. O., 1993, ApJ, 418, 82 
%%%%\bibitem{}Mihos J. C., Hernquist L., 1994, ApJ, 437, 611
%%\bibitem{}Miller M. C., Colbert E. J. M., 2004, International Journal of Modern Physics D, 13, 1
%\bibitem{}Miller M. C., Hamilton D. P., 2002, MNRAS, 330, 232
%%%\bibitem{}Miller B. W., Hodge P., 1994, ApJ, 427, 656
%%%\bibitem{}Miller B. W., Hodge P., 1996, ApJ, 458, 467
%%%\bibitem{}Moustakas J., Kennicutt R. C., Jr., 2006a, ApJ, 651, 155
%%%\bibitem{}Moustakas J., Kennicutt R. C., Jr., 2006b, ApJS, 164, 81
\bibitem{}Muijres L., Vink J. S., de Koter A., Hirschi R., Langer N., Yoon S.-C., 2012, A\&{}A, 546, 42
%%%\bibitem{}Mu$\tilde{\rm n}$oz-Mateos J. C., Gil de Paz A., Boissier S., Zamorano J., Jarrett T., Gallego J., Madore B. F., 2007, ApJ, 658, 1006
%%\bibitem{}Mushotzky R., 2004, Progress of Theoretical Physics Supplement, 155, 27
%\bibitem{}Naiman J. P., Ramirez-Ruiz E., Lin D. N. C., 2012, arXiv:1206.5002
%\bibitem{}Nanni A., Bressan A., Marigo P., Girardi L., Danese L., 2013, MNRAS, submitted
%\bibitem{}Narayan R., McClintock J. E., 2005, ApJ, 623, 1017
\bibitem{}Nelemans G., Yungelson L. R., Portegies Zwart S. F., Verbunt F., 2001, A\&{}A, 365, 491 
%%%\bibitem{}Oliva E., Marconi A., Moorwood A. F. M., 1999, A\&{}A, 342, 87O
%\bibitem{}Orosz  J. A., 2003, in  van der Hucht  K. A., Herrero  A., Esteban, C., eds, Proc. IAU Symp. 212, {\it A Massive Star Odyssey: From Main Sequence to Supernova}
%\bibitem{}Orosz  J. A., et al., 2007, Nature, 449, 872 
%\bibitem{}\"Ozel F., Psaltis D., Narayan R., McClintock J. E., 2010, ApJ, 725, 1918
%%%\bibitem{}Pagel B. E. J., Simonson E. A., Terlevich R. J., Edmunds M. G., 1992, MNRAS, 255, 325
%%\bibitem{}Pakull M. W., Mirioni L., 2002, astro-ph/0202488
%%%\bibitem{}Pastoriza M. G., Dottori H. A., Terlevich E., Terlevich R., Diaz A. I., 1993, MNRAS, 260, 177
%%\bibitem{}Patruno A., Colpi M., Faulkner A., Possenti A., 2005, MNRAS, 364, 344
%\bibitem{}Patruno A., Zampieri L., 2008, MNRAS, 386, 543	
%\bibitem{}Pattabiraman Bh., Umbreit S., Liao W.-K., Choudhary A., Kalogera V., Memik G., Rasio F. A., 2012, submitted to ApJ Supplements, arXiv:1206.5878
\bibitem{}Pauldrach A. W. A., Vanbeveren D., Hoffmann T. L. 2012,A\&{}A, 538, 75
\bibitem{}Pfalzner S., 2009, A\&{}A, 498, L37 
%%%\bibitem{}Pilyugin L. S., 2001a, A\&{}A, 369, 594 [P01]
%%%\bibitem{}Pilyugin L. S., 2001b, A\&{}A, 374, 412
%%%\bibitem{}Pilyugin L. S., Izotova I. Yu., Sholudchenko Yu. S., 2008, Kinematics and Physics of Celestial Bodies, 24, 192 
%%%\bibitem{}Pilyugin L. S., Moll\'a M., Ferrini F., V\'ilchez J. M., 2002, A\&{}A, 383, 14
%%%\bibitem{}Pilyugin L. S., Thuan T. X., 2005, ApJ, 631, 231 [PT05]
%%%\bibitem{}Pilyugin L. S., Thuan T. X., 2007, ApJ, 669, 299
%\bibitem{}Pilyugin L. S., V\'ilchez J. M., Contini T., 2004, A\&{}A, 425, 849 
%%%%\bibitem{}Pilyugin L. S., Thuan T. X., 2007, ApJ, 669, 299
%%%\bibitem{}Pilyugin L. S., Thuan T. X., V\'ilchez J. M., 2006, MNRAS, 367, 1139
%\bibitem{}Podsiadlowski Ph., Rappaport S., Pfahl E. D., 2002, ApJ, 565, 1107
%\bibitem{}Podsiadlowski Ph., Rappaport S., Han Z., 2003, MNRAS, 341, 385 
\bibitem{}Portegies Zwart S. F., Verbunt F., 1996, A\&{}A, 309, 179
%\bibitem{}Portegies Zwart S. F., Verbunt F., Ergma E., 1997, A\&{}A, 321, 207 
%%\bibitem{}Portegies Zwart S. F., McMillan S. L. W., 2000, ApJ, 528, L17
\bibitem{}Portegies Zwart S. F., McMillan S. L. W., Hut P., Makino J., 2001, MNRAS, 321, 199
\bibitem{}Portegies Zwart S. F., McMillan S. L. W., 2002, ApJ, 576, 899
\bibitem{}Portegies Zwart S. F., 2004, arXiv:astro-ph/0406550, Lecture note to appear in ``Joint Evolution of Black Holes and Galaxies'' of the Series in High Energy Physics, Cosmology and Gravitation. IOP Publishing, Bristol and Philadelphia, 2005, eds M. Colpi, V.Gorini, F.Haardt and U.Moschella
%%\bibitem{}Portegies Zwart S. F., Dewi J., Maccarone T. 2004, MNRAS, 355, 413
\bibitem{}Portegies Zwart S. F., McMillan S. L. W., Makino J., 2007, MNRAS, 374, 95
\bibitem{}Portegies Zwart S. F., McMillan S. L. W.,, Gieles M., 2010, ARA\&{}A, 48, 431
\bibitem{}Portinari L., Chiosi C., Bressan A., 1998, A\&{}A, 334, 505
%%\bibitem{}Poutanen J., Lipunova G., Fabrika S., Butkevich A. G., Abolmasov P., 2007, MNRAS, 377, 1187
%\bibitem{}Prestwich A. H., Kilgard R., Crowther P. A., Carpano S., Pollock A. M. T., Zezas A., Saar S. H., Roberts T. P., Ward M. J., 2007, ApJ, 669, L21
%%%\bibitem{}Ptak A., Colbert E., 2004, ApJ, 606, 291
\bibitem{}Puzia Th. H., Kissler-Patig M., Brodie J. P., Huchra J. P., 1999, AJ, 118, 2734
%\bibitem{}Quinlan G. D.,  1996, NewAstronomy, 1, 35
%%%\bibitem{}Ranalli P., Comastri A., Setti G., 2003, A\&{}A, 399, 39
%\bibitem{}Rangelov B., Prestwich A. H., Chandar R., 2011, ApJ, 741, 86
%\bibitem{}Rappaport S. A., Podsiadlowski Ph., Pfahl E., 2005, MNRAS, 356, 401
%%\bibitem{}Rasio F. A., Heggie D. C., 1995, ApJ, 445, L133
%%%\bibitem{}Read A. M., 2003, MNRAS, 342, 715
\bibitem{}Richards E. E., Lang C. C., Trombley Ch., Figer D. F., 2012, AJ, 144, 89
%%\bibitem{}Roberts T. P., 2007, Astrophysics and Space Science, 311, 203
%%%%\bibitem{}Ryder S. D., 1993, Ph.D. Thesis, Aust. Natl. Univ. Canberra, 149
%%%\bibitem{}Ryder S. D., Dopita M. A., 1994, ApJ, 430, 142
%%\bibitem{}Salpeter E. E., 1955, ApJ, 121, 161
%%%\bibitem{}Schmitt H. R., Calzetti D., Armus L., Giavalisco M., Heckman T. M., Kennicutt R. C. Jr., Leitherer C., Meurer G. R., 2006, ApJ, 643, 173
\bibitem{}Schulman R. D., Glebbeek V., Sills A., 2012, MNRAS, 420, 651
%\bibitem{}Sepinsky J., Kalogera V., Belczynski K., 2005, ApJ, 621L, 37
\bibitem{}Sigurdsson S., Phinney E. S., 1993, ApJ, 415, 631
%\bibitem{}Silverman J. M., Filippenko A. V., 2008, ApJ, 678, L17
\bibitem{}Sippel A. C., Hurley J. R., Madrid J. P., Harris W. E., 2012, MNRAS, 427, 167 
%%%\bibitem{}Skillman E. D., Kennicutt R. C., Hodge P. W., 1989, ApJ, 347, 875
%\bibitem{}Smartt S. J., 2009, ARA\&{}A, 47, 63
%%%\bibitem{}Smith B. J., Struck C., Nowak M. A., 2005, AJ, 129, 1350
%%%\bibitem{}Smith D. A., Wilson A. S., 2001, ApJ, 557, 180
%%\bibitem{}Socrates A., Davis S. W., 2006, ApJ, 651, 1049
%\bibitem{}Sollima A., Carballo-Bello J. A., Beccari G., Ferraro F. R., Fusi Pecci F., Lanzoni B., 2010, MNRAS, 401, 577 
%\bibitem{}Soria R., Cropper M., Pakull M., Mushotzky R., Wu K., 2005, MNRAS, 356, 12
%\bibitem{}Soria R., 2006, in Populations of High Energy Sources in Galaxies, Edited by  Meurs E. J. A.; Fabbiano G.. Cambridge: Cambridge University Press, 473
%\bibitem{}Soria R., Kuntz K. D., Winkler P. F., Blair W. P., Long K. S., Plucinsky P. P., Whitmore B. C., 2012, ApJ, 750, 152
%%\bibitem{}Soria R., Cropper M., Pakull M., Mushotzky R., Wu K., 2005, MNRAS, 356, 12
%%%\bibitem{}Spinoglio L., Andreani P., Malkan M. A., 2002, ApJ, 572, 105
\bibitem{}Spitzer L., 1987 Dynamical evolution of globular clusters, Princeton University Press
%%%\bibitem{}Stevens I. R., Forbes D. A., Norris R. P., 2002, MNRAS, 335, 1079
%%\bibitem{}Stobbart A.-M., Roberts T. P., Wilms J., 2006, MNRAS, 368, 397
%\bibitem{}Stodolkiewicz J. S., 1982, Acta Astronomica, 32, 63
%\bibitem{}Stodolkiewicz J. S., 1986, Acta Astronomica, 36, 19
\bibitem{}Strader J., Fabbiano G., Luo B., Kim D.-W., Brodie J. P., Fragos T., Gallagher J. S., Kalogera V., King A., Zezas A., 2012, ApJ, 760, 87
%%%\bibitem{}Strateva I. V., Komossa S., 2009a, ApJ, 692, 443
%%%\bibitem{}Strateva I. V., Komossa S., 2009b, MNRAS, submitted, arXiv:0903.1548 
%%%\bibitem{}Strohmayer T. E., Mushotzky R. F., 2003, ApJ, 586L, 61
%%%\bibitem{}Strohmayer T. E., Mushotzky R. F., Winter L., Soria R., Uttley P., Cropper M., 2007, ApJ, 660, 580
%%%%\bibitem{}Struck-Marcell C., Higdon J. L., 1993, ApJ, 411, 108
%%%%\bibitem{}Struck C., Appleton P. N., Borne K. D. Lucas R. A., 1996, AJ, 112, 1868
%%%\bibitem{}Swartz D. A., Ghosh K. K., Tennant A. F., Wu K., 2004, ApJS, 154, 519
%%\bibitem{}Swartz D. A., Soria R., Tennant A. F., 2008, ApJ, 684, 282
%\bibitem{}Swartz D. A., Tennant A. F., Soria R., 2009, ApJ, 703, 159
%\bibitem{}Swartz D. A., 2010, Galaxy Wars: Stellar Populations and Star Formation in Interacting Galaxies ASP Conference Series Vol. 423. Edited by Smith B., Bastian N., Higdon S. J. U., and Higdon J. L.. San Francisco: Astronomical Society of the Pacific, p.277
%%%%%\bibitem{}Talent D. L., 1980, Ph.D. Thesis, 99
%%\bibitem{}Tauris T. M., van den Heuvel E., 2006, In: Compact stellar X-ray sources. Edited by Lewin W., van der Klis M. Cambridge Astrophysics Series, No. 39. Cambridge, UK: Cambridge University Press, 623
%\bibitem{}Thompson T. A., Quataert E., Burrows A., 2005, ApJ, 620, 861
%%%\bibitem{}Trinchieri G., Fabbiano G., Bandiera R., 1989, ApJ, 342, 759 
%%%\bibitem{}Tsch\"oke D., Hensler G., Junkes N., 2000,A\&{}A, 360, 447
\bibitem{}Tumlinson J., Shull J. M., 2000, ApJ, 528, L65
%\bibitem{}Vanbeveren D., 2009, New Astronomy Reviews, 53, 27 
%%\bibitem{}van der Marel R. P., 2004, in Ho L., ed., Coevolution of Black Holes and Galaxies, Cambridge Univ. Press, p. 37 
%%Published by Cambridge University Press, as part of the Carnegie Observatories Astrophysics Series. 
\bibitem{}van der Hucht K. A., 1991, in `Wolf-Rayet Stars and Interrelations with Other Massive Stars in Galaxies': Proceedings of the 143rd Symposium of the International Astronomical Union. Edited by Karel A. van der Hucht and Bambang Hidayat. International Astronomical Union. Symposium no. 143, Kluwer Academic Publishers, Dordrecht, p.19
%%%\bibitem{}Vigroux L., Stasi\'nska G., Comte G., 1987, A\&{}A, 172, 15
\bibitem{}Vink J. S., de Koter A., Lamers H. J. G. L. M. , 2001, A\&{}A, 369, 574
\bibitem{}Vink J. S., de Koter A., 2005, A\&{}A, 442, 587
%\bibitem{}Voss R., Nielsen M. T. B., Nelemans G., Fraser M., Smartt S. J., 2011, MNRAS, 418, L124
%%%\bibitem{}Walter F., et al., 2007, ApJ, 661, 102
%%%\bibitem{}Wang Q. D., Whitaker K. E., Williams R., 2005, MNRAS, 362, 1065
%%%\bibitem{}Wang Q., Wu X., 1992, ApJS, 78,391
%%%%%\bibitem{}Weaver K. A., Heckman T. M., Dahlem M., 2000, ApJ, 534, 684 
%%%\bibitem{}Webster B. L., Smith M. G., 1983, MNRAS, 204, 743
%\bibitem{}Wheeler J. C., Akiyama S., Williams P. T., 2005, Ap\&{}SS, 298, 3
%%%\bibitem{}Winter L. M., Mushotzky R. F., Reynolds C. S., 2006, ApJ, 649, 730
%%\bibitem{}Winter L. M., Mushotzky R. F., Reynolds C. S., 2007, ApJ, 655, 163
%%%%\bibitem{}Wolter A., Trinchieri G., Iovino A., 1999, A\&{}A, 342, 41
%%%\bibitem{}Wolter A., Trinchieri G., 2004, A\&{}A, 426, 787
%%%%\bibitem{}Wolter A., Trinchieri G., Colpi M., 2006, MNRAS, 373, 1627
\bibitem{}Woodley K. A., G\'omez M., 2010, Publications of the Astronomical Society of Australia, 27, 379
%%%\bibitem{}Woosley S. E., Weaver T. A., 1986, ARA\&{}A, 24, 205
%\bibitem{}Woosley S. E., Heger A., Weaver T. A., 2002, Reviews of Modern Physics, 74, 1015
%\bibitem{}Woosley S. E., Bloom J. S., 2006, ARA\&{}A, 44, 507
%%\bibitem{}Zampieri L., Mucciarelli P., Falomo R., Kaaret P., Di Stefano R., Turolla R., Chieregato M., Treves A., 2004, ApJ, 603, 523
%\bibitem{}Zampieri L., Roberts T., 2009, MNRAS, 400, 677
%\bibitem{}Zezas A., Fabbiano G., Rots A. H., Murray S. S., 2002, ApJ, 577, 710
%\bibitem{}Zuo Z.-Y., Li X.-D., 2010, MNRAS, 405, 2768
\end{thebibliography}
\end{document}